\documentclass[manuscript]{emulateapj}
\usepackage{graphics}
\usepackage{amssymb}
\usepackage{color}
\usepackage{enumerate}

\begin{document}

\title{The two-phase formation history of spiral galaxies traced by the cosmic evolution of the bar fraction}

\author{Katarina Kraljic}
\affil{Laboratoire AIM Paris-Saclay, CEA/IRFU/SAp, CNRS/INSU, Universit\'e Paris Diderot, 91191 Gif-sur-Yvette Cedex, France.}

\author{Fr\'ed\'eric Bournaud}
\affil{Laboratoire AIM Paris-Saclay, CEA/IRFU/SAp, CNRS/INSU, Universit\'e Paris Diderot, 91191 Gif-sur-Yvette Cedex, France.}

\and

\author{Marie Martig}
\affil{Centre for Astrophysics and Supercomputing, Swinburne University of Technology, PO Box 218, Hawthorn, VIC 3122, Australia.}

\begin{abstract}
We study the evolution of galactic bars and the link with disk and spheroid formation in a sample of zoom-in cosmological simulations. Our simulation sample focuses 
on galaxies with present-day stellar masses in the $10^{10-11}$\,M$_{\odot}$ range, in field and loose group environments, with a broad variety of mass growth histories. 
In our models, bars are almost absent from the progenitors of present-day spirals at $z>1.5$, and they remain rare and generally too weak to be observable down to $z  \approx  1$. After this characteristic epoch,
 the fractions of observable and strong bars raise rapidly, bars being present in 80\% of spiral galaxies and easily observable in two thirds of these at $z \leq 0.5$. 
This is quantitatively consistent with the redshift evolution of the observed bar fraction, although the latter is presently known up to $z \approx0.8$ because of 
band-shifting and resolution effects. Our models hence predict that the decrease in the bar fraction with increasing redshift should continue with a fraction of 
observable bars not larger than 10-15\% in disk galaxies at $z>1$. Our models also predict later bar formation in lower-mass galaxies, in agreement with existing data. 
We find that the characteristic epoch of bar formation, namely redshift $z  \approx 0.8-1$ in the studied mass range, corresponds to the epoch at which today's spirals
 acquire their disk-dominated morphology. At higher redshift, disks tend to be rapidly destroyed by mergers and gravitational instabilities and rarely develop significant 
bars. We hence suggest that the bar formation epoch corresponds to the transition between an early ``violent'' phase of spiral galaxy formation at $z\geq 1$ and a late 
``secular'' phase at $z \leq 0.8$. In the secular phase, the presence of bars substantially contributes to the growth of the (pseudo-)bulge, but the bulge mass budget 
remains statistically dominated by the contribution of mergers, interactions and disk instabilities at high redshift. 
Early bars at $z>1$ are often short-lived, while most of the bars formed at $z\leq1$ persist down to $z=0$, late cosmological gas infall being necessary to maintain some of them.
\end{abstract}

%%%%%%%%%%%%%%%%%%%%%%%%%%%%%%%%%%%%%%%%%%%%%%%%%%

\section{Introduction}

Bars are one of the most frequently and easily quantified substructures in spiral galaxies, and are hence often used as a tracer of galaxy evolution. Most spiral 
galaxies today contain a central bar, although with largely variable amplitudes \citep{block02,whyte02}. Spiral arms are equally common in optical light, but are much 
weaker in the near-infrared light that more closely traces the stellar mass distribution, and the strength of bars is generally easier to quantify independently of the 
imaging sensitivity.

Both the formation of bars and their time evolution are connected to the baryonic and dark matter properties of their host galaxies, and their mass assembly history. 
Bars form spontaneously in stellar disks that are sufficiently massive and dynamically cold to be gravitationally unstable, with typical Toomre 
stability parameters $Q \approx 1.5-2.0$ \citep{toomre63, combes-sanders, gerin, combes-elmegreen}, and can also be amplified by dynamical friction on the dark matter 
halo \citep{2002ApJ...569L..83A}. The gaseous content can also trigger bars: gas helps to form outer spiral arms, which can remove angular momentum from the inner regions and 
strengthen a bar seed \citep{BCS05}.

Once formed, bars evolve through the exchange of angular momentum with the dark matter halos \citep{1985MNRAS.213..451W,deb-sel2000}, as well as with stellar and gaseous disks \citep{1993A&A...268...65F,BC02,BCS05}. This can reinforce bars when they lose angular momentum, but can also weaken and destroy them in the opposite case \citep{BCS05}. 
Bars can also be weakened or destroyed by the growth of central mass concentrations \citep{norman,pfenniger,hasan}, however central concentrations with low-enough mass 
and/or low-enough mass growth rates could have little effect on real bars \citep{2005MNRAS.363..496A}. If bars are formed in conditions where they are intrinsically 
short-lived, sufficient accretion of external gas onto the disk could enable their survival or re-formation \citep{BC02}. In addition, galaxy interactions could in 
theory trigger bar (re-)formation \citep{gerin,1998ApJ...499..149M,2004MNRAS.347..220B}, although observations do not show a clear environmental dependence of the bar 
in disk galaxies \citep{vdbergh,2009A&A...495..491A,2011MNRAS.410L..18B,2012ApJ...745..125L}.

Hence, the fraction of barred galaxies, and the redshift evolution of this fraction, are fundamental {\em tracers} of the evolution history of galaxies: this indicates 
when disks became sufficiently massive and self-gravitating to be bar-unstable, and whether the conditions for bars being long-lived or re-formed were met. Furthermore, 
bars can directly drive structural evolution of their host galaxies. They trigger radial gas flows and may provide gas to nuclear disks and central black holes 
\citep{2000ApJ...529...93K,2002ApJ...567...97L,2004ApJ...607..103L}. They can also thicken through vertical resonances, leading to the formation of pseudo-bulges -- i.e.
, bulges with relatively low concentration and substantial residual rotation \citep[e.g.,][]{1999AJ....118..126B,kormendy-kennicutt,martinez-v06}.

\medskip

In the nearby Universe, the bar fraction in disk galaxies is very high. Depending on classification techniques, the fraction of strong bars in the optical light is at 
least $50\%$ \citep{2008ApJ...675.1194B}. Optical classifications reveal roughly one third of strongly barred galaxies, one third of weakly or moderately barred 
galaxies, and one third of optically unbarred galaxies \citep{1991trcb.book.....D}. In the near-infrared, where weak bars are not obscured by dust and more easily 
distinguished from spiral arms, the bar fraction is at least $80\%$ \citep{2000AJ....119..536E,block02,whyte02,2007ApJ...657..790M}.

The first searches for bars at redshift $z>0.5$ found a very low bar fraction \citep[e.g.,][]{1999MNRAS.308..569A}, possibly because of small number statistics.
Their work also illustrated the difficulties to identify bars at high redshift: the observed optical light traces the ultraviolet emission, in which bars are harder 
to detect, even locally. Near-infrared data revealed a number of barred galaxies at $z \geq 0.7$  \citep{2003ApJ...592L..13S,2004ApJ...612..191E,2004ApJ...615L.105J}. 
The first sample large enough to robustly quantify the redshift evolution of the bar fraction without being affected by resolution and band-shifting bias up to 
$z \approx 0.8$, was studied by \cite{2008ApJ...675.1141S} in the COSMOS field. 
These observations indicate that {\em the bar fraction drops by a factor of about three from $z=0$ to $z=0.8$}. This result holds both for all observable bars and for 
strong bars separately, and using either visual classifications or quantitative estimates of the bar strength. \citet{2008ApJ...675.1141S} also found a downsizing-like 
behaviour for bar formation, i.e. more massive galaxies tend to get barred at higher redshifts. This trend can explain why previous studies, such as \cite{2004ApJ...615L.105J},
using shallower data targeted to more massive systems, observed higher bar fractions -- but still consistent with a declining bar fraction \citep[see also][]{2004ApJ...612..191E}.

\medskip

In this paper, we study the evolution of bars in a sample of cosmological zoom-in simulations of 33 galaxies with present-day stellar masses ranging from $1\times 10^{10}$ to
$2\times 10^{11}$\,M$_{\sun}$, in field and loose group environments. The simulation technique and structural evolution of these galaxies (bulge and disk fractions, 
angular momentum evolution) were presented in \citet{martig12}.
The paper is organized as follows. In Section~\ref{Sec:simu_analysis} we present our simulations and methods for the identification of bars and morphological analysis.
In Section~\ref{Sec:redshift_evolution} we analyze the redshift evolution of the bar fraction in the whole sample and in disk-dominated galaxies\footnote{most galaxies in this sample are disk-dominated spirals at $z=0$, but a larger fraction is spheroid-dominated at $z>1$}, using quantitative measurements of the bar strength. 
Our main result, the emergence of bars along the cosmic time that traces the epoch of thin disk formation subsequent to the growth of spheroids and thick disks, is presented in Section~\ref{Sec:bars_vs_thin_disks}.
Section~\ref{Sec:bar_lifetime} studies the lifetime of bars and its dependence on external gas accretion. In Section~\ref{Sec:bars_vs_bulge},
we quantify the contribution of bars in the late growth of bulges, comparing to unbarred galaxies.
Finally, we discuss and summarize our results in Sections~\ref{Sec:discussion} and~\ref{Sec:summary}, respectively.

%%%%%%%%%%%%%%%%%%%%%%%%%%%%%%%%%%%%%%%%%%%%%%%%%%

\section{Simulations and Analysis}
\label{Sec:simu_analysis}
\subsection{Simulation Sample}

The simulation sample studied here was presented in \citet{martig12}. It comprises 33 field and loose group galaxies modeled at 150~pc resolution from redshift 5
down to redshift zero with present-day stellar masses ranging from $1\times 10^{10}$ to $2\times 10^{11}$\,M$_{\sun}$.

The zoom-in simulation technique is fully described in \cite{martig09, martig12}. Dark matter haloes are selected in a large volume, dark matter-only simulation, 
performed using $\Lambda$-cold dark matter cosmology with the Adaptive Mesh Refinement code RAMSES \citep{teyssier02}. The zoom-in simulation for each selected halo is performed by 
recording and replicating the mass inflow through a spherical boundary at the final virial radius of the selected halo. The boundary conditions used for the zoom-in 
simulation hence replicate all minor and major mergers as well as diffuse infall, as imposed by the initial cosmological simulation.

Our zoom-in simulations use a spatial resolution of 150~pc, and a mass resolution of $10^{4-5}$\,M$_{\sun}$ ($1.5\times10^4$\,M$_{\sun}$ for gas, 
$7.5\times10^4$\,M$_{\sun}$ for stars and $3\times10^5$\,M$_{\sun}$ for dark matter particles). The code used to model gravity for gas, stars and dark matter 
is the particle-mesh code described in \cite{BC02,BC03}.
Interstellar gas dynamics is modeled with a sticky-particle scheme \citep[studied in the context of bar evolution in][]{BCS05}, 
which has the drawback of neglecting thermal pressure (especially in hot halos, not crucial in the studied mass range), but the advantage of modeling the properties of 
supersonic turbulent pressure in cold gas phases -- which is the physically dominant pressure term in the star-forming interstellar medium \citep[e.g.][]{burkert06} and is hardly 
modeled by other hydrodynamic techniques {\em unless} resolutions of $1-10$~pc and low thermal cooling floors are reached 
\citep{2010ApJ...720L.149T, bournaud11,2011MNRAS.417..950H}. Star formation is computed with a Schmidt-Kennicutt law \citep{1998ApJ...498..541K} with an 
exponent of $1.5$ and an efficiency of $2\%$. The star formation threshold is set at $0.03$\,M$_{\sun}$pc$^{-3}$. Energy feedback from supernovae explosions using a kinetic scheme as well as the continuous gas mass-loss from stars \citep{2001A&A...376...85J,2010ApJ...714L.275M} are included.
The $z=5$ seed for the central galaxy and the incoming companions are implemented with arbitrary disk+bulge models; tests have shown that the seed properties at $z=5$ 
have no substantial impact on the structural evolution from $z=2$ out to $z=0$, because rapid evolution and mass growth at $z>3$ washes out the initial assumptions.

In our study, we focus on the redshift range from $z=2$ to $z=0$. Detailed description of analyzed sample is presented in \citet{martig12}. Here, we 
recall typical masses and star formation rates (SFR) at different redshifts. At $z=0$, typical mass and SFR of simulated galaxies are $4-5\times10^{10}$\,M$_{\sun}$ 
and $3$\,M$_{\sun}$yr$^{-1}$, at $z=1$, $2\times10^{10}$\,M$_{\sun}$ and $5$\,M$_{\sun}$yr$^{-1}$ and at $z=2$, $2-3\times10^9$\,M$_{\sun}$ and $10$\,M$_{\sun}$yr$^{-1}$, 
respectively. Hence, a broad range of mass growth histories is covered by the sample (see \citealt{martig12} for details).

\subsection{Bar analysis}
\label{Sec:bar_analysis}

The method used to determine the presence, length and strength of a bar is based on the azimuthal spectral analysis of surface density profiles. This is obtained by
considering the stellar surface density of each galaxy in polar coordinates, decomposed into Fourier components in the form:
\begin{equation}
\label{eq:Fourier}
 \Sigma(r,\theta)= \Sigma_0(r)+\sum_m A_m(r) \cos(m\theta - \Phi_m(r)),
\end{equation}
where $\Sigma(r,\theta)$ is the stellar surface density, $\theta$ the azimuthal angle given in rotating frame in which the bar is fixed and $r$ the radial distance. 
$A_m$ and $\Phi_m$ are the associated Fourier amplitude and phase, respectively. $\Sigma_0(r)$ is the azimuthally-averaged profile of the stellar surface density. 
The analysis is performed on a face-on projection: the spin axis of the whole stellar content of the galaxy is used to define the corresponding line of sight.
The center of mass of the stars within the central 10 kpc is taken to be the center of the galaxy for the Fourier analysis.

A typical signature of the presence of bar is the prominence of even components, especially $m=2$, within the bar region. The identification of bar is possible by 
studying the phase $\Phi_2 (r)$ which is constant with radius in the bar region, as opposed to a two-armed spiral mode ($\Phi_m$ varies linearly with radius for an 
$m$-armed spiral mode).

After careful examination of the whole sample at different redshifts, we decide that a bar is present if $\Phi_2(r)$ is constant to within $\pm 5\,^{\circ}$ around the 
median value over a large-enough region, hereafter the ``bar region''. This bar region should start at radii between 900~pc and 1500~pc, and cover a radial range of at 
least 1500~pc. We search the starting point of the bar region at $r \geq 900$~pc because central asymmetries or off-centering in the central resolution elements can 
cause variations of $\Phi_2$ at smaller radii even for visually barred systems, while starting the search at 900~pc is found to never exclude systems that are visually 
identified as barred. We stop the search at 1500~pc as no bar identified visually starts its $\Phi_2 \simeq$~constant region at larger radii. We impose the bar region 
($\Phi_2$ constant to within $10\,^{\circ}$) to cover a radial range of at least 1500~pc because this selects all bars identified visually {\em and} this excludes 
spiral arms, as the latter typically have a variation of $\Phi_2$ of a few tens of degrees over a few kpc. The choice of cutoff in the bar length is given mainly by the resolution. We require the bar to lie within at least three resolution bins used in Fourier analysis. 
Moreover, typical lengths of significant bars are in general $\geq2$ kpc \citep{2008ApJ...675.1194B}. Smaller bars are usually either nuclear bars (which are not the 
focus of this study) or weak ones. We could possibly miss some of these short and weak bars, but this does not have any impact on our conclusions on bars that would be 
strong enough to be observed at $z>0$.

Once a bar region meeting to the previous criteria is found, the galaxy is classified as barred and the bar strength $S$ is measured following the definition proposed, 
for instance, by \cite{aguerri1998}: 
\begin{equation}
\label{eq:strength}
 S \equiv r_{bar}^{-1}\int_0^{r_{bar}} \!  \frac{A_2}{A_0} \, \mathrm{d}r,
\end{equation}
where $r_{bar}$ is the outer radius of the bar region.

\begin{figure*}[here]
\centering
\includegraphics[width=0.8\textwidth]{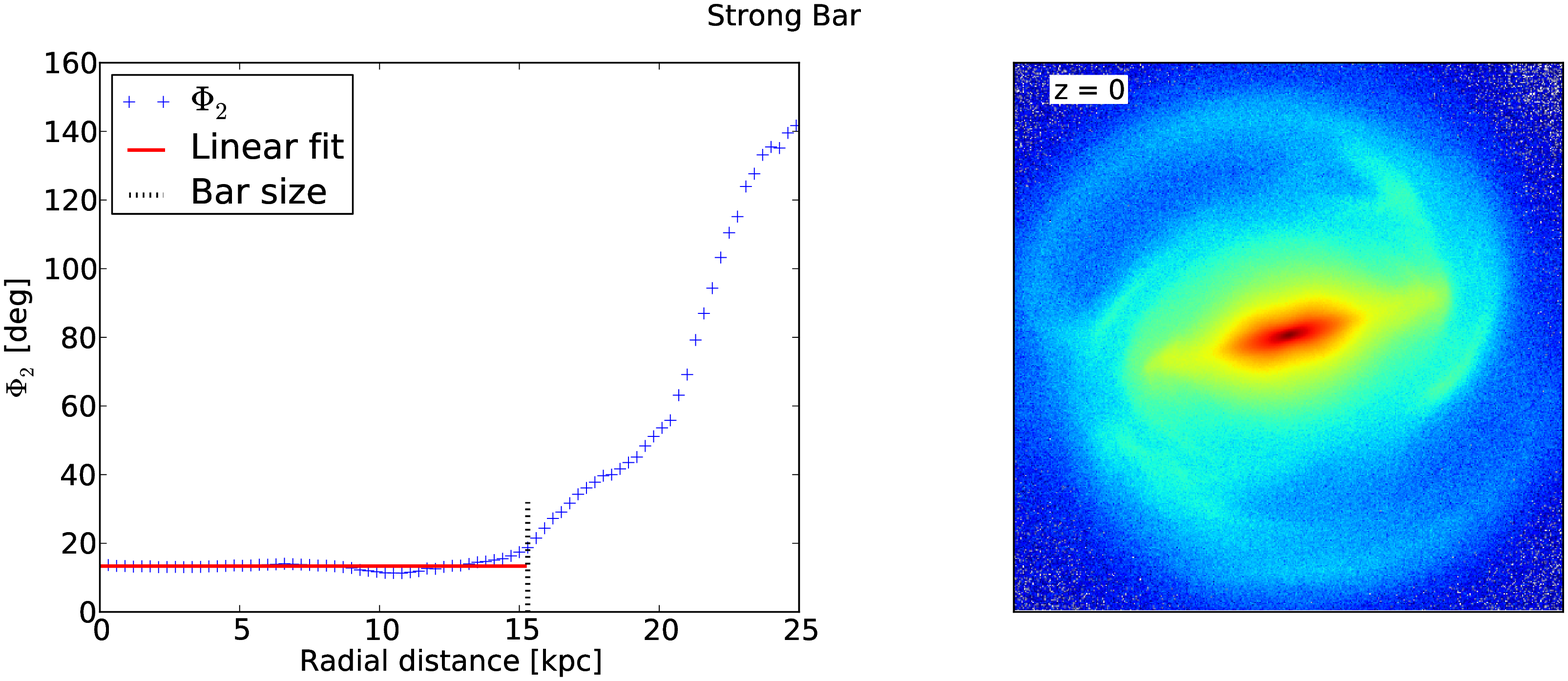}\\
\includegraphics[width=0.8\textwidth]{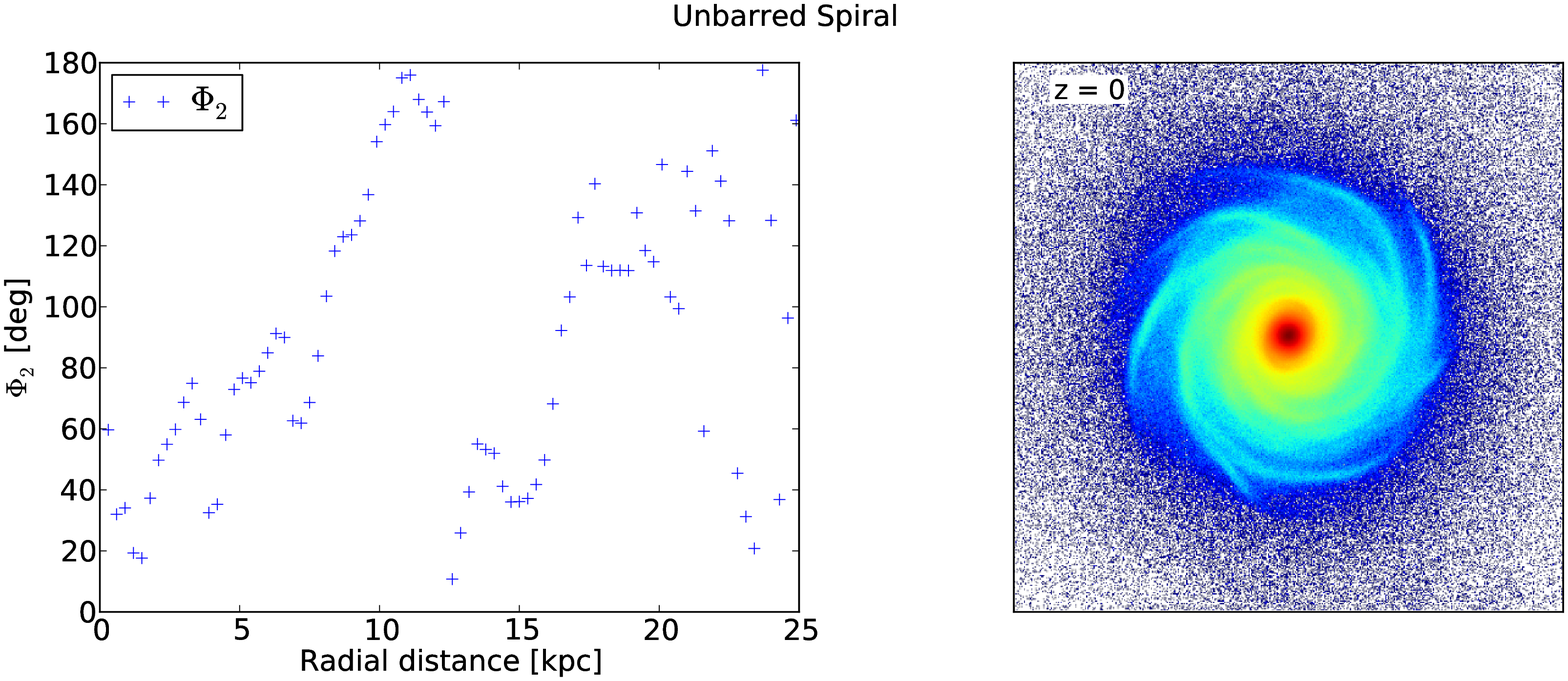}
\caption{\label{fig:lin_fit}
Examples of the fitting method results (left panels) and the corresponding surface density maps (right panels) viewed face-on ($50 \times 50$ kpc$^2$)
for two galaxies in our sample. The top panel shows a (strongly) barred galaxy with the region where the phase $\Phi_2$ is almost constant identified as the bar region,
while the bottom panel corresponds to an unbarred galaxy for which no region of constant $\Phi_2$ is detected. The color coding scale of surface density
map is logarithmic going from $\sim 10^{0.5}$\,M$_{\sun}$pc$^{-2}$ for dark blue to $\sim 10^4$\,M$_{\sun}$pc$^{-2}$ for dark red.}
\end{figure*}

Figure~\ref{fig:lin_fit} shows two examples of barred and unbarred galaxies and the corresponding radial profiles of $\Phi_2$. With this technique, the automated 
identification of barred galaxies is in agreement with all the cases that have been visually examined. We classify bars into strong bars and weak bars by considering 
the strength $S$ (Eq.~\ref{eq:strength}). In the following, we will distinguish:
\begin{itemize}
\item {\em all detected bars}, including very weak and short bars,
\item {\em observable bars}, with a strength $S \geq 0.2$, which is the typical detection limit used in observations up to high redshift
(e.g., \citealt{2008ApJ...675.1141S}) -- even at $z=0$, weaker bars may be confused with spiral arms unless the $\Phi_2$ phase can be probed accurately
with very deep imaging (e.g., \citealt{block02}),
\item  and {\em strong bars} with $S \geq 0.3$ unless specified otherwise.
\end{itemize}

\begin{figure*}[here]
\centering
\includegraphics[width=0.8\textwidth]{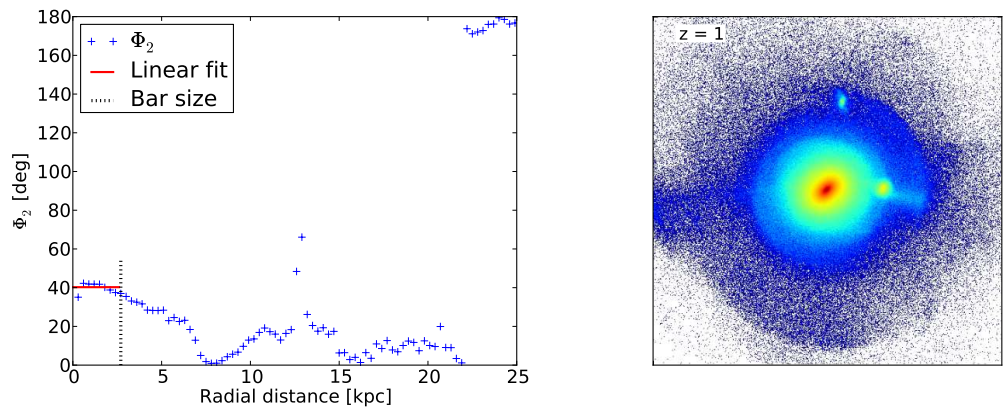}
\caption{\label{fig:lin_fit3} Example of a possible ``fake'' bar. A region of constant $\Phi_2$ is identified at $r\leq2.7$ kpc (left panel), corresponding to the detection of a short bar.
However, the face-on density map (right panel, $50 \times 50$ kpc$^2$) suggests that the identified region could instead be the core of an elliptical, i.e., a ``fake'' bar, which is confirmed by the examination of edge-on projections performed in Figure~\ref{fig:3images_2}.
The color coding of the projected density map is the same as in Figure~\ref{fig:lin_fit}.}
\end{figure*}

\begin{figure*}[t]
\centering
\includegraphics[width=0.8\textwidth]{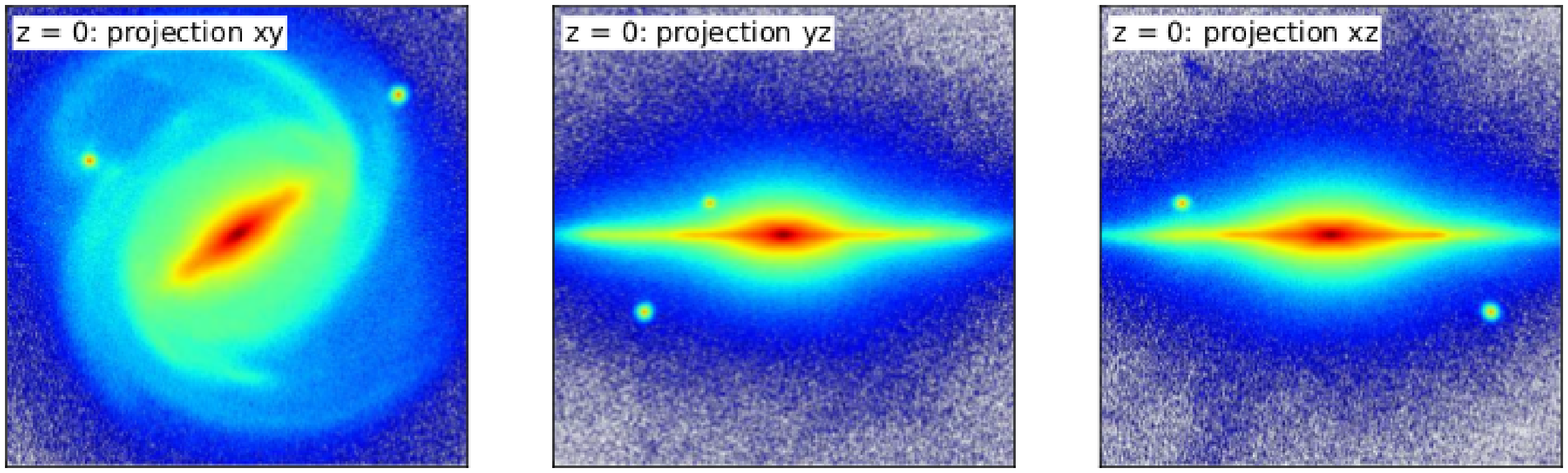}\\
\vskip 1cm
\includegraphics[width=0.8\textwidth]{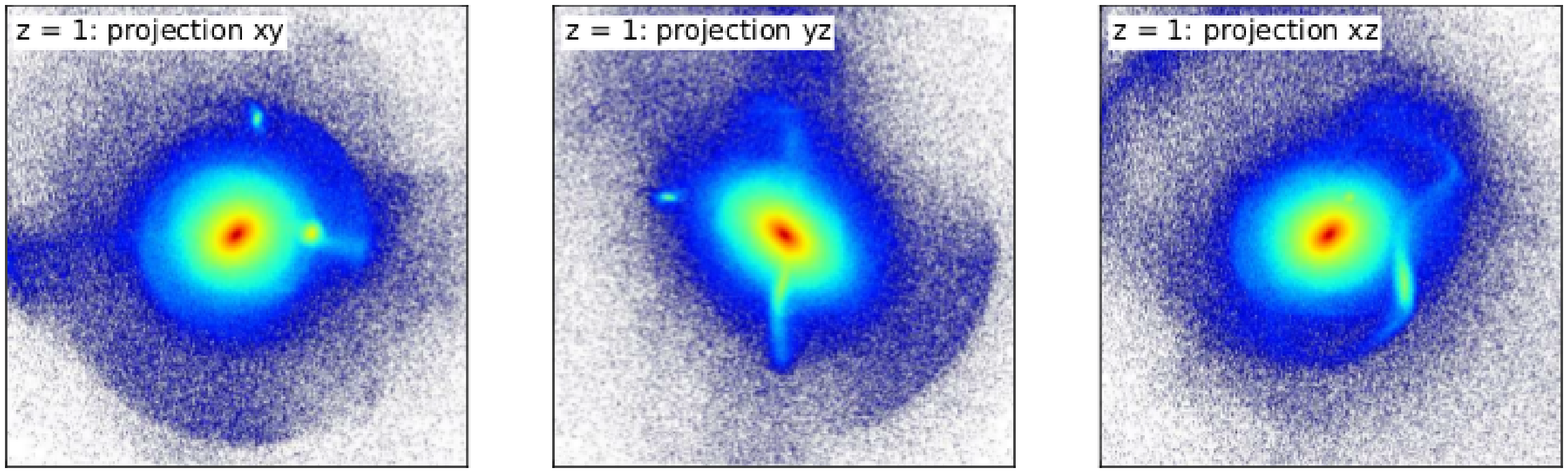}
\caption{\label{fig:3images_2} Examples of surface density maps (50$\times$50\,kpc$^2$) in three different projections for two simulated galaxies. The corresponding redshift
is $z=0$ and $z=1$ for galaxies in top and bottom panels, respectively. If a potential bar is identified in the face-on projection, the two orthogonal edge-on projections are used to
discriminate real bars from triaxial cores in spheroids or ``fake'' bars. The color coding of the projected density maps is the same as in Figure~\ref{fig:lin_fit}.}
\end{figure*}

At this stage, the identification of bars can still be confused with
flattened early type galaxies, especially for weak bars. A spheroid-dominated galaxy,
in the face-on projection defined by the stellar spin axis, can have flattened isophotes in its central regions,
which could be identified as a ``bar'' with our definition, while this
corresponds to a triaxial (part of) stellar spheroid rather than a real bar: such an example is shown on Figure~\ref{fig:lin_fit3}.
This occurs only with weak bars, and there are few spheroid-dominated galaxies in our sample at $z=0$, so the problem is unimportant there, but such cases are 
somewhat more frequent at increasing redshift. We thus need to distinguish such ``fake'' bars from real (if weak) bars. We find that examining two edge-on projections 
of the stellar mass density unambiguously solves the problem for our whole sample. Real weak bars are found in galaxies with a massive disk component, hence edge-on 
projections of the stellar density are substantially flat. ``Fake'' bars are found in galaxies that are spheroid-dominated (at least in the central few kpc) and the 
edge-on projections are quite round (see Figure~\ref{fig:3images_2} for an illustration). We thus decide that galaxies for which the strength $S$ of the $m=2$
mode in two orthogonal edge-on projections\footnote{In these projections the
$m=2$ mode strength is not a bar strength, but traces the presence of an edge-on disky component in the stellar mass distribution.} is greater or equal to 0.3 
correspond to real weak bars in disky stellar systems. Other cases are almost round in all projections ($S<0.3$ in all projections) and are considered to be 
spheroid-dominated galaxies with ``fake'' bars, i.e. moderately flattened central regions.

These ``fake'' bars are rare at low redshift (about $6\%$ of all galaxies) and represent less than $15\%$ of all galaxies at $z>1.5$. Visual inspection of the results 
showed that at most two thirds of the ``fake'' bars are successfully removed by the technique described above at any redshift, so any remaining contamination of the 
bar fraction would be quite minor. Moreover, most of the results studied in this paper relate to the fraction of bars among disk-dominated galaxies at various 
redshifts (based on a S\'{e}rsic index measurement~\citep{1963BAAA....6...41S}, see Section~\ref{Sec:morphology}), which is not contaminated by ``fake'' bars. Indeed, these ``fake'' bars are 
flattened spheroid-dominated galaxies, which are naturally removed from the S\'{e}rsic index-selected sample of disk-dominated galaxies.

It has been shown by \citet{BCS05} that the $m=2$ modes are not strongly affected by the dissipation parameters used in sticky particle codes. We have performed additional checks to estimate the impact of the centering of the galaxy on the result of the Fourier analysis. Changing the in-plane coordinates of the center used for the Fourier decomposition by as much as 300~pc, we do not find any
significant change in the detection and strength of bars classified as observable (as defined above). Some galaxies identified as unbarred can appear weakly barred, 
and weak bars may not be detected anymore, when the center used for the Fourier decomposition is moved by as much as 300~pc. Such ambiguous cases are quite rare for 
centering offsets smaller than 300~pc. The overall rate of misclassification due to the galaxy off-centering is not larger than 15\% and affects mostly bars that would not be observable at high redshift.

\subsection{Morphology analysis: Identifying disk-dominated and bulge-dominated galaxies}
\label{Sec:morphology}

In order to distinguish disk-dominated galaxies from earlier-type ones, we fit the radial profile of the stellar mass surface density with a S\'{e}rsic profile of form:
\begin{equation}
 \Sigma(r) = a_0 \exp\left[- a_1
\left(\frac{r}{r_0}\right)^{\frac{1}{n}} \right],
\end{equation}
where $a_0$, $a_1$ are normalization constants, $n$ is the S\'{e}rsic index, $r$ is the radius and $r_0$ is the scale-length. Galaxies with a S\'{e}rsic index $n\leq2$ are classified as
disk-dominated. The fitted range is $r_{50} \leq r  \leq2\times r_{90}$ for unbarred galaxies and $bar \leq r\leq 2\times r_{90}$ (or $bar \leq r\leq 2.5\times i_{25}$
if the length of the bar is shorter than $r_{90}$) for barred galaxies, where $r_{50}$ and $r_{90}$ are radii containing $50\%$ and $90\%$ of mass, respectively, $bar$ 
is the corrected length of the bar (length of the bar - $900$~pc) and $i_{25}$ the $25^{th}$ isophote. The scale-length $r_0$ is set to the value of $r_{50}$ for each 
galaxy. The ranges for our set of simulated galaxies are chosen to produce a satisfactory match to the observed morphology of these galaxies on one side and 
staying relatively simple with comprehensible physical interpretation on the other side. Indeed, the present classification is reproducible in observations and it 
agrees with the classification based on a decomposition of disk and bulge components in the 6-D phase space by \citet{martig12} for the vast majority of the sample. The galaxies 
presently classified as disk-dominated have bulge-to-total mass fraction below 35\%. 

Note that the identification of a disk component to reject ``fake bars'' (Section~\ref{Sec:bar_analysis}) was based on a different criterion. There we aimed at probing the presence of a 
substantial disky component hosting the bar, without requiring it to dominate the stellar mass distribution, and we used edge-on projections for this. Here we use the 
S\'{e}rsic index of the face-on projection to probe whether the whole galaxy is dominated by a disk component typically twice more massive (at least) than the bulge.
As the S\'{e}rsic profile is fitted to the stellar mass density, it is not sensitive to the nuclear concentrations of star formation. Nevertheless, we checked the consistency of the identification with the bulge-to-total ratios obtained by \citet{martig12}.

\begin{figure}[here]
\centering
\resizebox{\columnwidth}{!}{\includegraphics{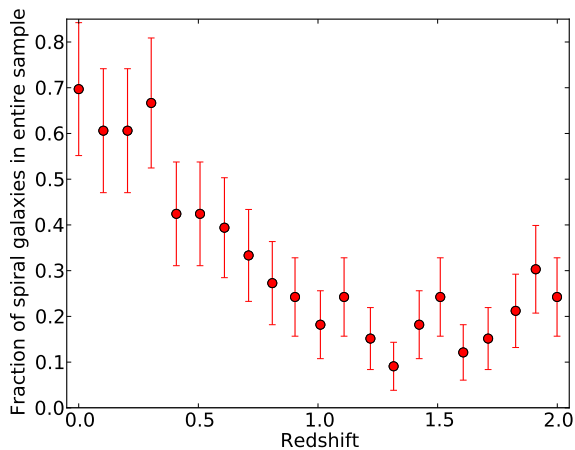}}
\caption{\label{fig:frac_spirals}Evolution with redshift of the fraction of galaxies classified as spirals according to their S\'{e}rsic index as defined in the text.
At $z=0$ about two thirds of galaxies are found to be disk-dominated while for $z>1$ this fraction is $\sim0.2$ with $\pm1 \sigma$ fluctuations.
Here and in all following figures, the error bars on measured fractions are Poissonian, the average error bar being attributed to zero values, unless stated differently.}
\end{figure}

Figure~\ref{fig:frac_spirals} shows the evolution with redshift of the fraction of disk-dominated galaxies within all 33 simulated galaxies. $70\%$ of galaxies are 
found to be disk-dominated at $z=0$, and their bulge fraction is 23\% (from the \citealt{martig12} decomposition), consistent with classifying them as ``spiral galaxies''. The 
remaining $30\%$ are galaxies with an average bulge-to-total ratio of 0.43 and could be classified as early-type S0 galaxies. We find a decreasing fraction of 
disk-dominated galaxies as a function of redshift, down to a fraction of $20\%$ at $z=1$.

%%%%%%%%%%%%%%%%%%%%%%%%%%%%%%%%%%%%%%%%%%%%%%%%%%
\section{The redshift evolution of the bar fraction}
\label{Sec:redshift_evolution}
\subsection{Bar fraction history}
\label{Sec:bars_z_evolution}

\begin{figure}[here]
%\centering
\includegraphics[width=\columnwidth]{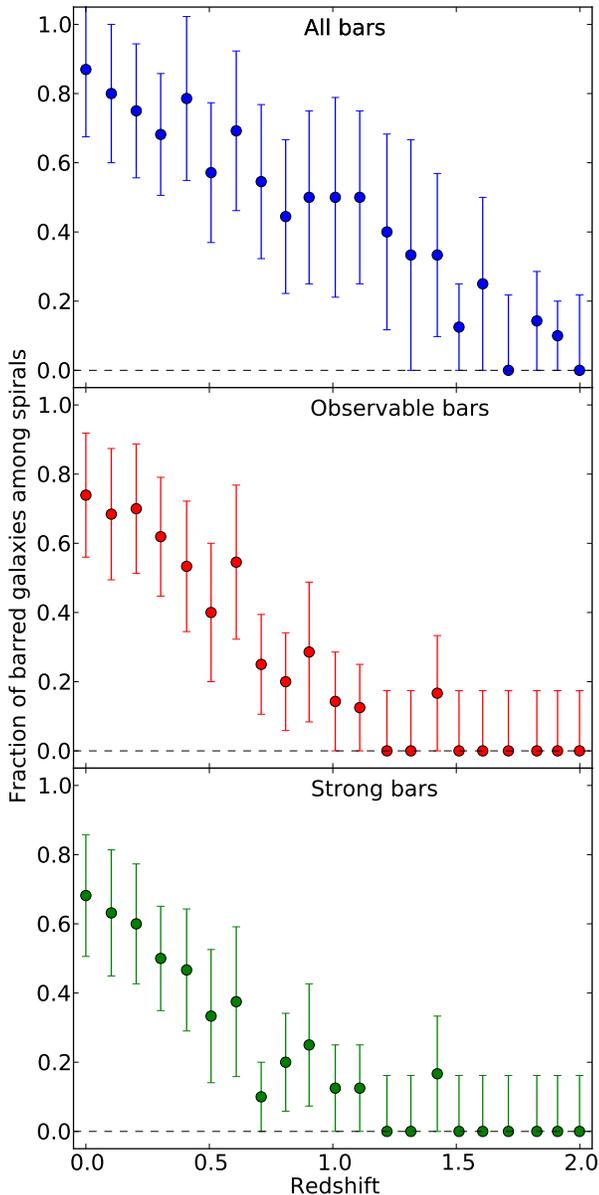}
\caption{\label{fig:all_bars_compare} Evolution with redshift of the total bar fraction including even weak bars (top panel), observable bar fraction (middle panel), 
and strong bar fraction (bottom panel) among spiral galaxies (i.e., galaxies with a S\'{e}rsic index $n<2$ in each redshift bin).}
\end{figure}

Figure~\ref{fig:all_bars_compare} shows the redshift evolution of the bar fraction among galaxies classified as spirals (i.e., disk-dominated). The top panel shows this
fraction including all values of the bar strength, while the middle and bottom panels show the redshift evolution of the bar fraction for observable bars 
(strength $S\geq0.2$) and strong bars ($S\geq0.3$), respectively.

The total bar fraction among spiral galaxies declines with increasing redshift. At redshift zero, 80-90\% of spiral galaxies contain a bar, while at $z \simeq1$ this 
fraction drops to about $50\%$, and to almost zero at $z \simeq 2$. Similarly, the observable bar fraction and strong bar fraction decline from about 70\% at $z=0$ to 
$10-20\%$ at $z \simeq 1$. Observable and strong bars are virtually absent at $z>1.5$.

\begin{figure}[here]
\centering
\resizebox{\columnwidth}{!}{\includegraphics{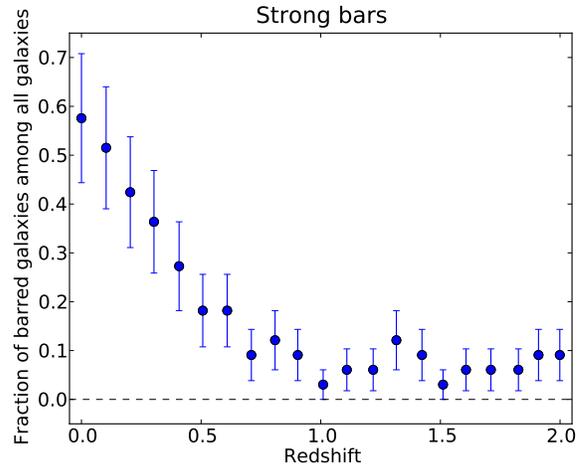}}
\caption{\label{fig:0.3_bars_all} Evolution with redshift of the fraction of barred galaxies among all galaxies, where only strong bars ($S\geq0.3$) are considered. 
Note that the bottom panel of Figure~\ref{fig:all_bars_compare} showed strong bars among galaxies classified as disk dominated (S\'{e}rsic index $n<2$).}
\end{figure}

The steady decline of the total, observable and strong bar fractions from $z=0$ to $z\sim1-2$ cannot be attributed to a decline of the fraction of disk-dominated 
galaxies, since the bar fractions mentioned above are measured among disk-dominated (spiral) galaxies. There is a decline in the fraction of disk galaxies with 
increasing redshift (Figure~\ref{fig:frac_spirals}), and as a result the bar fraction among all galaxies declines even more rapidly with redshift 
(from $\approx$\,$60\%$ at redshift $z=0$ to less than $5\%$ at $z=1$,  Figure~\ref{fig:0.3_bars_all}) but the decline of the bar fraction occurs even considering only 
disk-dominated galaxies.

\medskip

\begin{figure}[here]
\centering
\resizebox{\columnwidth}{!}{\includegraphics{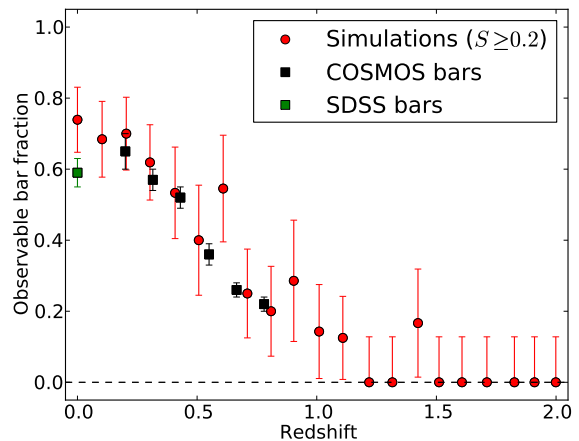}}
\caption{\label{fig:all_compare} Comparison of the redshift evolution of the bar fraction in spiral galaxies with COSMOS and SDSS data \citep{2008ApJ...675.1141S}. 
The lower limit on the bar strength of simulated galaxies is set to $0.2$, so that only observable bars are shown. The error bars are calculated as in 
\citet{2008ApJ...675.1141S}, assuming binomial statistics.}
\end{figure}

\begin{figure}[here]
\centering
\resizebox{\columnwidth}{!}{\includegraphics{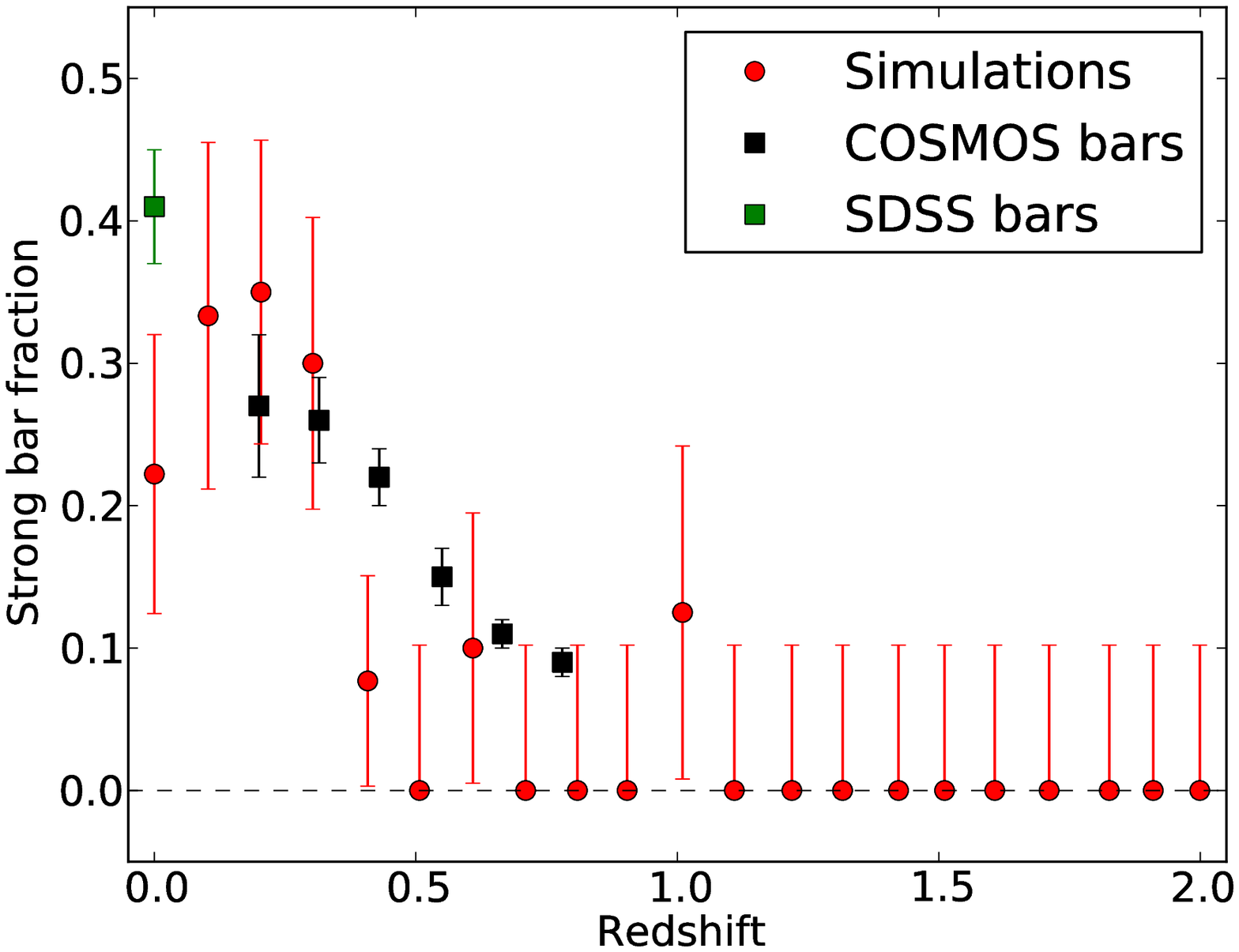}}
\caption{\label{fig:strong_compare}Comparison of the evolution with redshift of the fraction of strong bars (here considering strengths greater than 0.4) in spirals 
of COSMOS and SDSS data from \citet{2008ApJ...675.1141S}. The error bars are calculated as in \citet{2008ApJ...675.1141S}, assuming binomial statistics.}
\end{figure}

We compare in Figures~\ref{fig:all_compare} and~\ref{fig:strong_compare} our data to the observations of the $z = 0 - 0.8$ sample in COSMOS and the SDSS by 
\citet{2008ApJ...675.1141S}, which is the largest and highest-redshift dataset measuring the observed bar fraction to date. The redshift evolution of the bar fraction 
is in quantitative agreement between models and data, considering both the fraction of observable bars and of strong bars\footnote{defined in 
Figure~\ref{fig:strong_compare} as bars stronger than 0.4 ($S \geq 0.4$) for consistency with \citet{2008ApJ...675.1141S}}. Note that \citet{2008ApJ...675.1141S} 
used an ellipse-fitting technique to measure bar strengths, but such techniques give quantitative results quite consistent with Fourier decompositions 
\citep{block02,2002MNRAS.331..880L}, as well as with visual estimates by~\citet{2008ApJ...675.1141S} themselves. We also note that the fraction of bars that could 
be robustly observed at $z>0$, i.e. with a strength larger than 0.2, is lower than the total fraction of bars identified in the simulations including the weakest bars,
especially at redshift $z\sim 1$ (see top panel of Figure~\ref{fig:all_bars_compare} versus Figure~\ref{fig:all_compare}), but both fractions follow a similar 
evolution with redshift.

Observations of the bar fraction are typically limited to redshift $z \lesssim 0.8$ to date (with very rare $z \sim 1$ examples, \citealt{2004ApJ...612..191E}). 
Our models agree with the observed trends up to those redshifts, and suggest that the observed decline should continue at higher redshifts, observable bars being 
almost absent at $z \geq 1$.

\subsection{Dependence on stellar mass}
\label{Sec:mass}

\begin{figure}[here]
\centering
\includegraphics[width=\columnwidth]{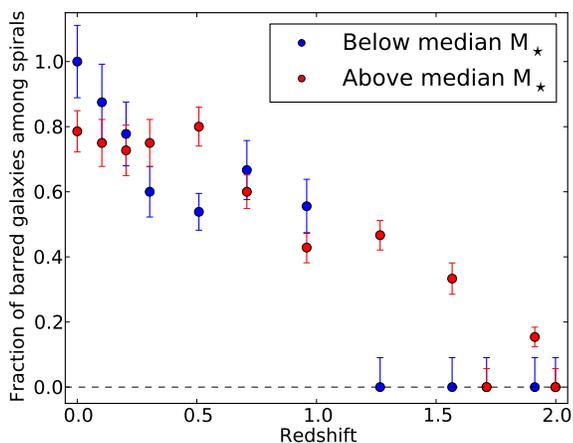}
\caption{\label{fig:mass_fraction_bars} Evolution with the redshift of the fraction of barred galaxies among spiral galaxies according to their mass. In each 
redshift bin, galaxies are classified as having a stellar mass above or below the median value, which is $4\times 10^9$\,M$_{\sun}$ at z=2, 
$2.5\times 10^{10}$\,M$_{\sun}$ at $z\sim1$ and $6\times 10^{10}$\,M$_{\sun}$ at z=0.}
\end{figure}

Figure~\ref{fig:mass_fraction_bars} shows the redshift evolution of the total bar fraction among spiral galaxies for low-mass and high-mass systems separately. We 
consider galaxies above and below the median mass of the sample in each redshift bin. The bin size was adjusted so that the Poissonian error in each bin does not 
exceed $20\%$.

We find that bars form later in lower-mass galaxies, in a ``downsizing''-like mode, which is consistent with the observations of~\citet{2008ApJ...675.1141S}. These 
authors proposed that merging activity, which is more common at high redshift, could affect low-mass systems more severely by heating them and thus delaying or 
preventing bar formation. However, mergers alone cannot account for the observed trend in our models. First, major mergers of
spiral galaxies often result in the formation of spheroid-dominated galaxies, even at high redshift \citep{bournaud11}. 
We find that the majority of disk galaxies are barred today, and bar formation is found in our sample even when considering only disk-dominated galaxies with low S\'{e}rsic indices, so the process should be independent on the occurrence of 
major mergers. Second, minor mergers and interactions can cause the destruction {\em or} formation of a bar,depending in particular on the orbital parameters 
\citep{gerin,2003MNRAS.341..343B,2004MNRAS.347..220B}. Since bars at high redshift ($z\gtrsim1$) are rare, any net effect of mergers/interactions should be bar 
formation in previously un-barred systems, rather than bar destruction in previously barred systems.

Over the entire sample, our results suggest that the epoch of bar formation is the typical epoch at which galaxies start to be dominated by a kinematically cold, 
thin stellar disk (see Section~\ref{Sec:bars_vs_thin_disks}). A possible explanation of the ``downsizing'' of bar formation could be that these modern spiral disks themselves form later in 
lower-mass galaxies. The continuation of rapid mass accretion onto lower-mass systems down to lower redshift could keep their disk violently unstable, with giant 
clumps and irregular structures rather than bars, as further discussed in Section \ref{Sec:discussion}. More massive spiral  galaxies seem to be largely in place 
and already disk-dominated at $z\sim1$ \citep[e.g.,][]{2007ApJS..172..434S}.

We also find that at intermediate redshift ($z\lesssim0.6$) the fraction of bars in low-mass galaxies continues to increase with decreasing $z$, while in high-mass 
systems, it stays roughly constant from $z\sim0.6$ to $z=0$ at a relatively high value of $\sim70\%$. At high redshift ($z>1$), the most massive disk galaxies do 
contain some bars (although relatively rare, and in general too weak to be easily observable), while lower mass disk galaxies are systematically unbarred.

\begin{figure}[here]
\centering
\includegraphics[width=\columnwidth]{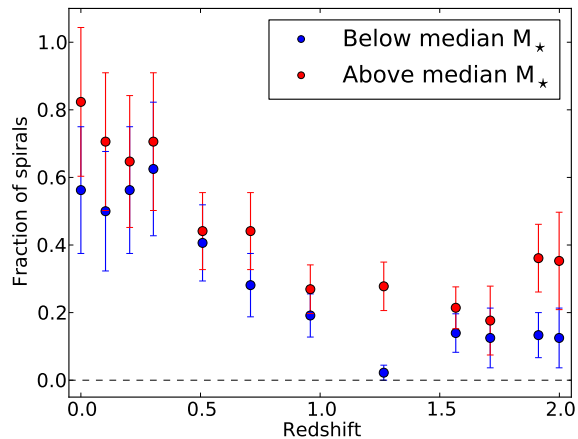}
\caption{\label{fig:mass_fraction_spirals} Evolution with the redshift of the fraction of spiral galaxies among all galaxies according to their mass. The binning is 
same as in Figure~\ref{fig:mass_fraction_bars}.  }
\end{figure}

%%%%%%%%%%%%%%%%%%%%%%%%%%%%%%%%%%%%%%%%%%%%%%%%%%%%
\section{When do bars emerge? The emergence of massive thin disks}
\label{Sec:bars_vs_thin_disks}

While the total bar fraction and strong bar fraction are both high at low redshift, they become very small or close to zero at $z \geq 1$. Most of the bars found at 
$z>1$ in the simulations are quite weak and would not even be considered with a bar strength threshold at $S=0.2$, hence they would be hardly observable at high 
redshift (see Figure~\ref{fig:all_compare}).

The redshift evolution of the fraction of disk-dominated\footnote{according to criterion defined in Section~\ref{Sec:morphology}} galaxies in our simulation sample 
was shown on Figure~\ref{fig:frac_spirals}. Surprisingly, it is quite similar to the redshift evolution of the fraction of strong bars or moderate (observable) bars 
(see Figure~\ref{fig:frac_spirals} compared to the bottom panel of Figure~\ref{fig:all_bars_compare} and Figure~\ref{fig:all_compare}). This suggests that the epoch of bar 
formation is also the epoch at which galaxies that are presently spirals start to be dominated by stellar disks. Indeed, at higher redshift, the progenitors of these 
present-day spirals are sometimes disk-dominated but are often spheroid-dominated, or have irregular morphologies such as clumpy disks (i.e., short lived unstable disks), 
or are interacting and merging systems (see \citealt{martig12} for a thorough study of the structural evolution of the simulated galaxies in the same sample).

Our study and that in \citet{martig12} indicate that these present-day spiral galaxies grow mostly through two phases. During an early ``violent'' phase at $z>1$, the system is 
quite often disturbed by important mergers (major mergers or multiple minor mergers) as well as violent disk instabilities (giant clumps). The morphology can evolve 
during this phase from disk-dominated to a spheroid without being stabilized toward the final disk-dominated structure. At $z<1$, present-day spirals evolve mostly 
through a ``secular'' phase when the morphology is generally stabilized to a disk-dominated structure (\citealt{martig12}). The bulge growth is then only slow and limited as 
important mergers and violent disk instabilities become rare, and almost absent after $z \simeq 0.7$.

As measured in \citet{martig12}, there is no correlation between the morphologies and the disk/bulge fractions between the early violent phase ($z>1$) and the present day.
While the systems enter the ``secular'' phase after $z \approx 1$ the bulge/disk fractions become more and more tightly correlated with the final 
$z=0$ values. Stars formed or accreted in the early violent phase end-up mostly in the thick disk, stellar halo and bulge at redshift zero. Stars formed in the secular 
phase at $z<1$ mostly grow the modern thin spiral disk, with a substantial but non dominant contribution to late bulge growth, especially in barred systems (see 
Section~\ref{Sec:bars_vs_bulge} for late secular bulge growth).

\medskip

The epoch of bar formation covers a relatively narrow range $z \simeq 0.7-1$ for the vast majority of galaxies in our sample, although the sample spans stellar 
masses of $1 \times 10^{10}$ to $2 \times 10^{11}$\,M$_{\sun}$. The comparison with the structural evolution shows that the (observable) bar fraction is constantly 
very low in the early violent phase of spiral galaxy formation, a phase during which disk-dominated systems can be present but are generally destroyed/reformed over 
short timescales, and uncorrelated with the final disk fraction. Once spiral galaxies enter their ``secular phase'' at $z\leq 1$, bars rapidly form and the bar 
fraction rapidly raises. This is the epoch at which the final thin spiral disk starts to form and dominate the stellar structure of these galaxies: the disk and bulge 
fractions can continue to evolve down to $z=0$ but in this phase the disk is not destroyed/reformed anymore, although the bar itself may sometimes be destroyed/reformed. 
At this point, the formation of a long-lived massive thin disk allows the bar to form in most of the progenitors of today's spirals.
 
The epoch of the emergence of bars thus traces the epoch at which modern thin disks are established and start to dominate the stellar mass distribution in present-day 
spirals. This is primarily the interpretation of the bar fraction in our simulation sample, but if the redshift evolution of the bar fraction agrees with observations 
(which is the case up to $z \simeq 0.8$, and needs to be probed at higher redshift) then it could suggest that the same interpretation applies to observed bars, namely 
that their emergence traces the typical epoch at which spiral galaxies establish their modern disk-dominated structure at $z=0.8-1$ (for present-day stellar masses 
in the $10^{10-11}$\,M$_{\sun}$ range). We will further discuss this hypothesis of a two-phase formation history traced by the emergence of bars, and illustrate with 
representative examples, in Section~\ref{Sec:discussion}.

%%%%%%%%%%%%%%%%%%%%%%%%%%%%%%%%%%%%%%%%%%%%%%%%%%
\section{Bar lifetime}
\label{Sec:bar_lifetime}

\begin{figure*}[t]
\includegraphics[width=\columnwidth]{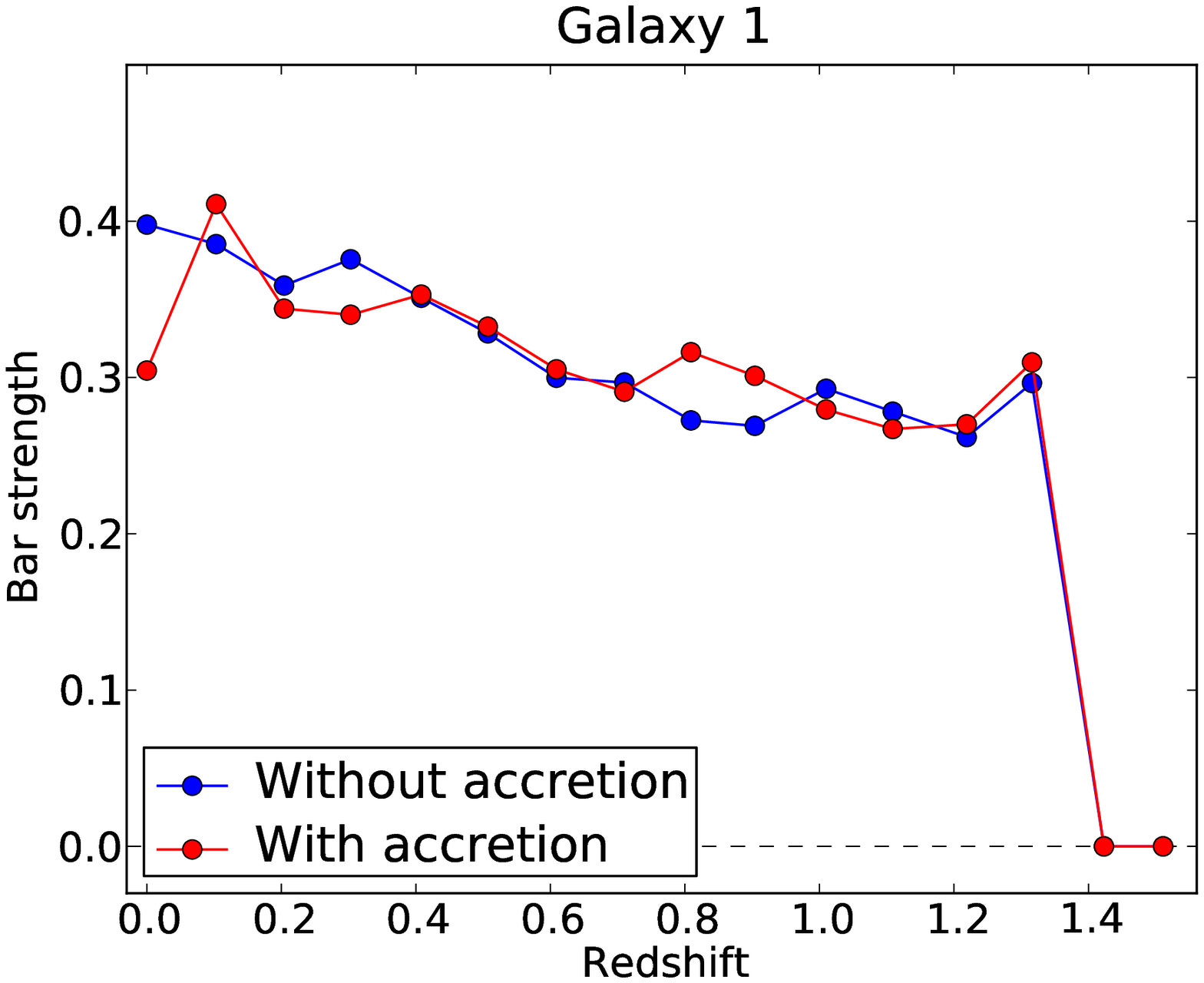}
\includegraphics[width=\columnwidth]{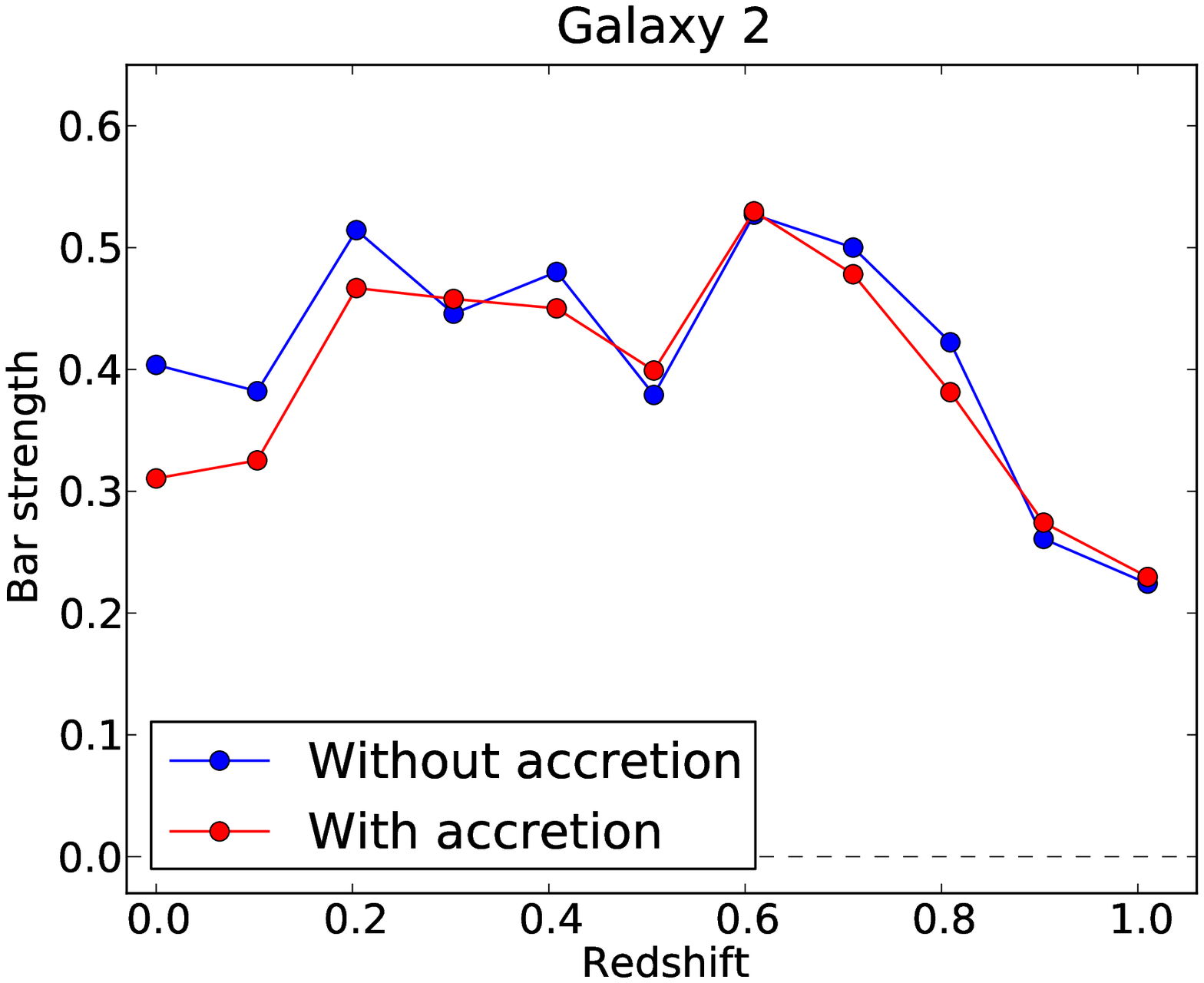}\\
\caption{\label{fig:similar_evolution} Examples of two galaxies with similar redshift evolution of bar strength with and without accretion. In both cases the two 
galaxies host long-lived strong ($S\gtrsim0.3$) bars that form at $z\gtrsim0.8$. The strength of bars does not evolve significantly. The external accretion hence 
does not seem to be needed to maintain a strong bar down to $z=0$. Here and in all following figures, the null value for the bar strength is to be interpreted as 
no bar detection rather than exact value obtained by Eq.~\ref{eq:strength}}.
\includegraphics[width=\columnwidth]{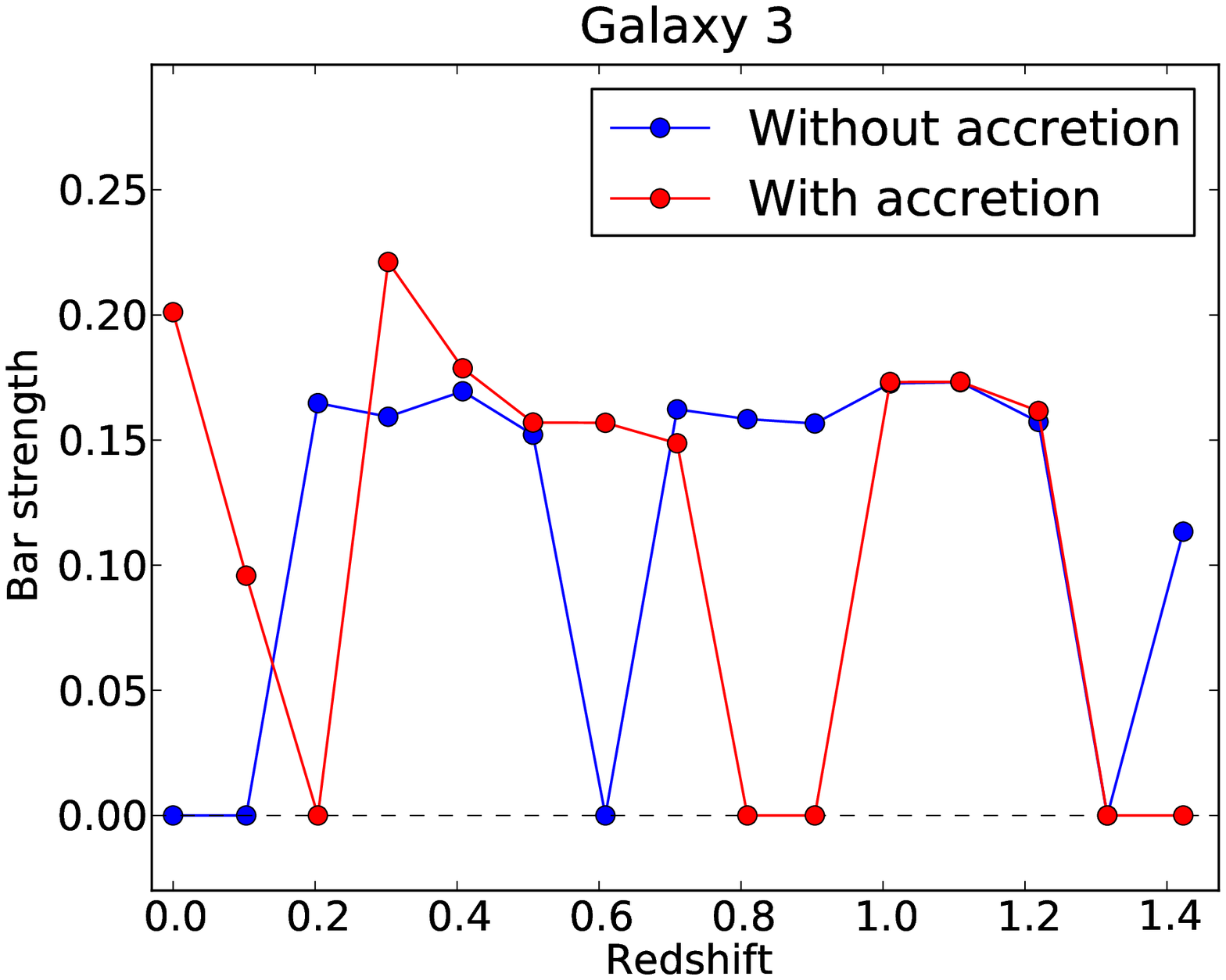}
\includegraphics[width=\columnwidth]{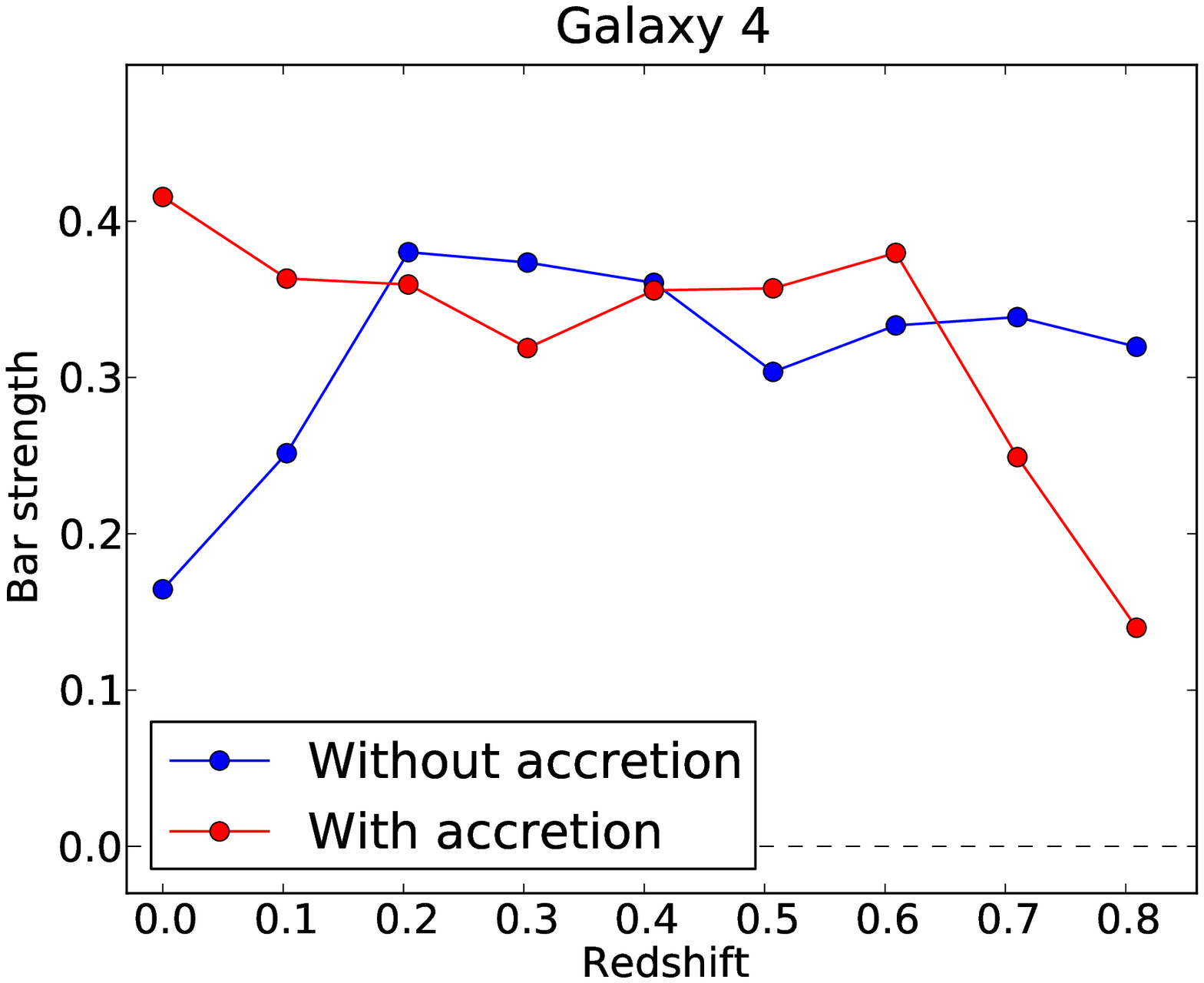}
\caption{\label{fig:different_evolution} Examples of two galaxies with different redshift evolution of bar strength with and without accretion. Galaxy 3 (left panel) 
hosts a weak bar that forms at $z>1$. Detailed redshift evolution of the bar strength in fiducial simulation is different than in the case without external accretion, 
but the overall behaviour is similar: bar undergoes several episodes of formation and dissolution. Galaxy 4 (right panel) hosts a strong bar from $z\sim0.6$ down
to $z\sim0$ in fiducial simulation, but weakens significantly from $z\sim0.2$ when external accretion is shut down.}
\end{figure*}

As shown in Section~\ref{Sec:bars_z_evolution}, bars start to steadily appear at redshift $z\sim0.8-1$. A closer examination of individual bars formed at this 
redshift reveals that their strength often remains roughly constant down to redshift $z\sim0$. Is this because the conditions are favorable for the strength being 
intrinsically constant, or would those bars tend to weaken or even be destroyed in complete isolation but are maintained by external infall as proposed by 
\cite{BC02}? We here examine the influence of the external accretion on the lifetime of a bar in realistic cosmological context (unlike the idealized accretion used 
by \cite{BC02}). We select four galaxies, hereafter labeled Galaxy~1 to Galaxy~4, with representative bar strength histories in the main sample, and we 
run new simulations without external accretion once their bar has formed.

Figure~\ref{fig:similar_evolution} shows the two first galaxies (Galaxy~1 and 2), for which the redshift evolution of the bar strength is very similar with and without 
accretion. The bar is formed at $z\sim1.3$ and $z\sim1$ in these two galaxies, respectively. In both galaxies the bar strength is roughly constant down to $z\sim0$ 
with and without accretion. These bars are intrinsically long-lived and their evolution is not significantly influenced by late mass accretion.

Figure~\ref{fig:different_evolution} shows the two other galaxies, for which external accretion has some influence on the bar strength evolution. In the 
fiducial simulation, Galaxy 3 (left panel) hosts a weak bar ($S\lesssim0.2$ at all redshifts), that dissolves and re-forms several times. Without external accretion, 
the bar evolution is different, but the overall trend is similar, with several episodes of formation and dissolution. In this galaxy the bar remains weak in all cases.

\begin{figure}[here]
\includegraphics[width=\columnwidth]{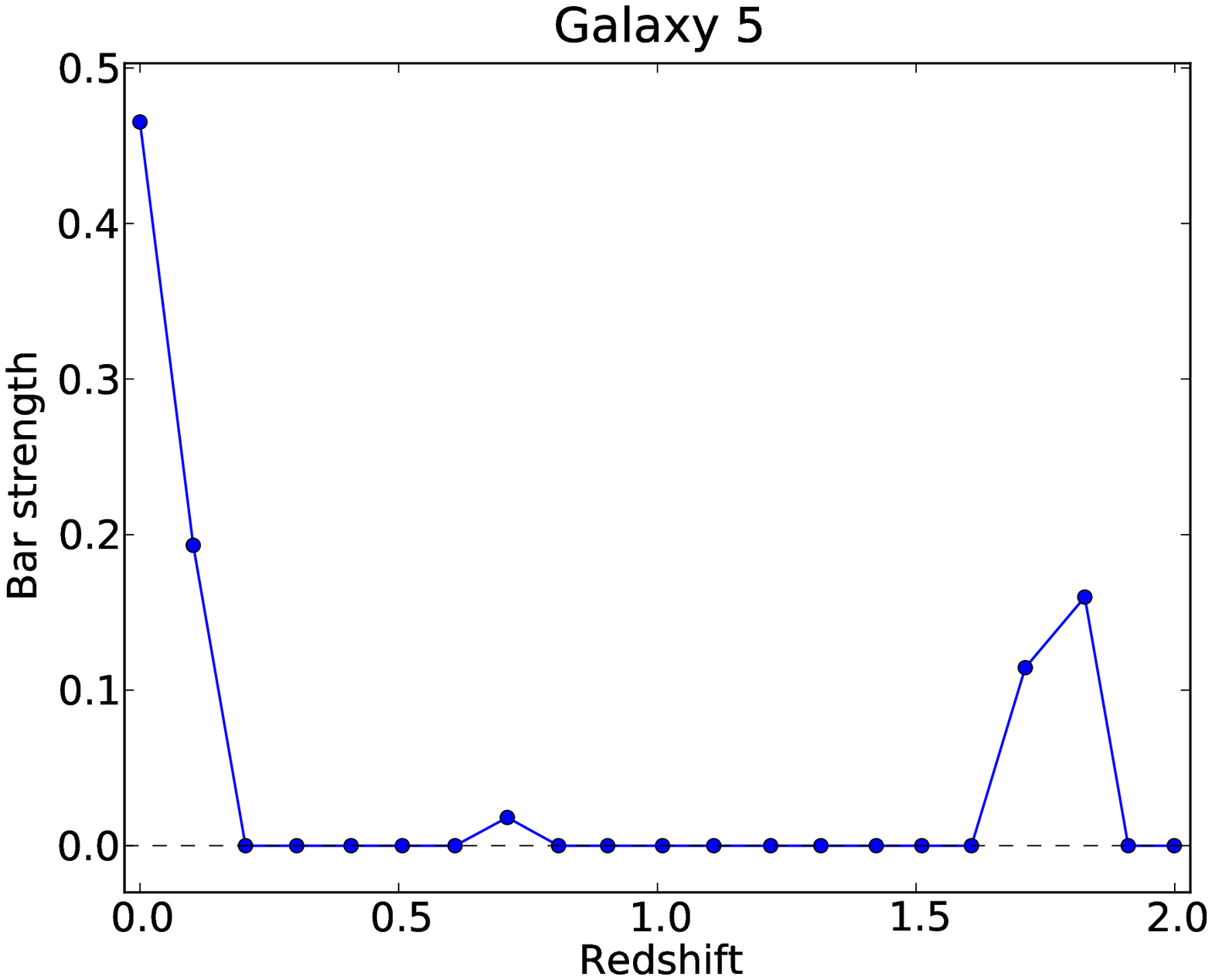}\\
\includegraphics[width=\columnwidth]{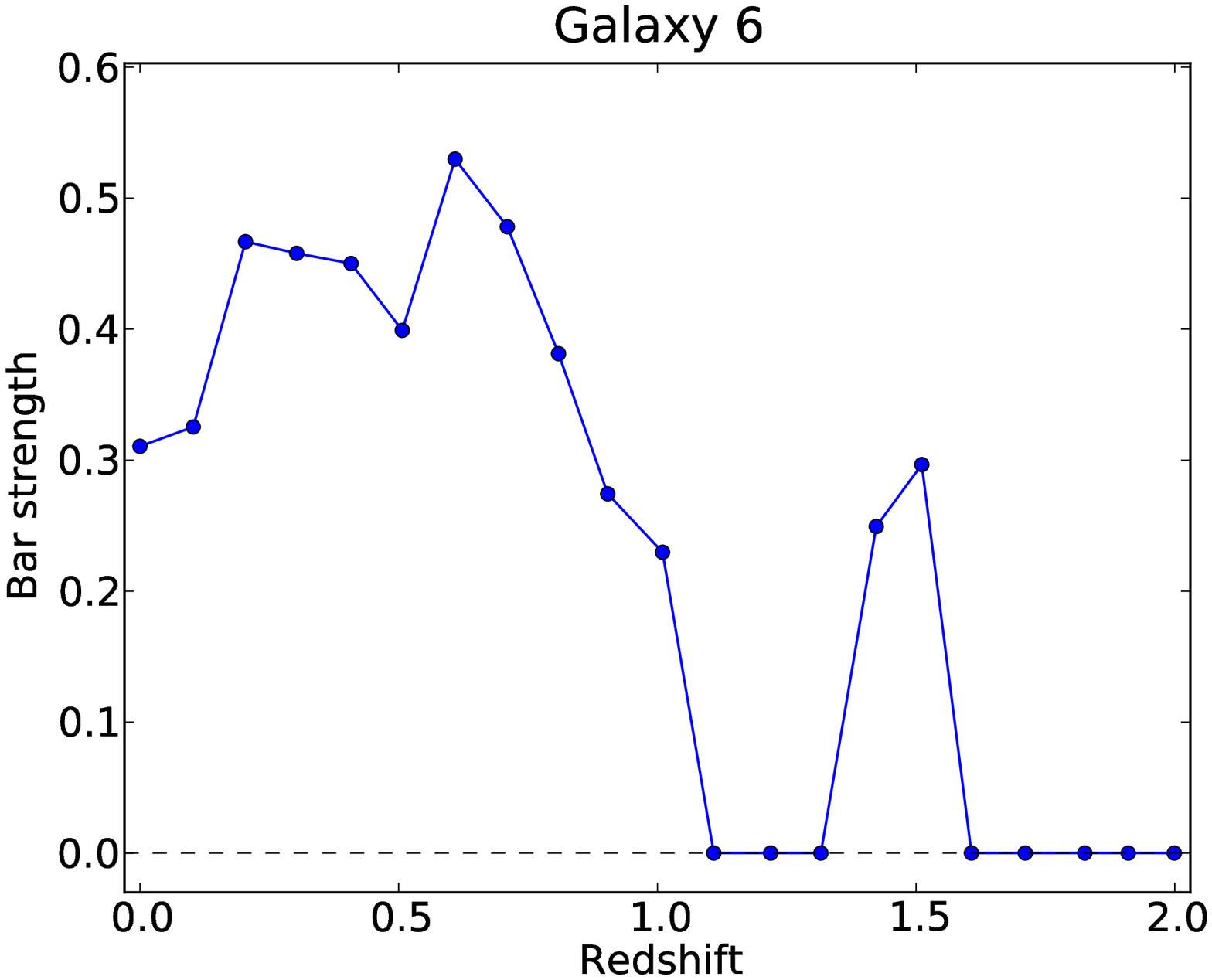}\\
\includegraphics[width=\columnwidth]{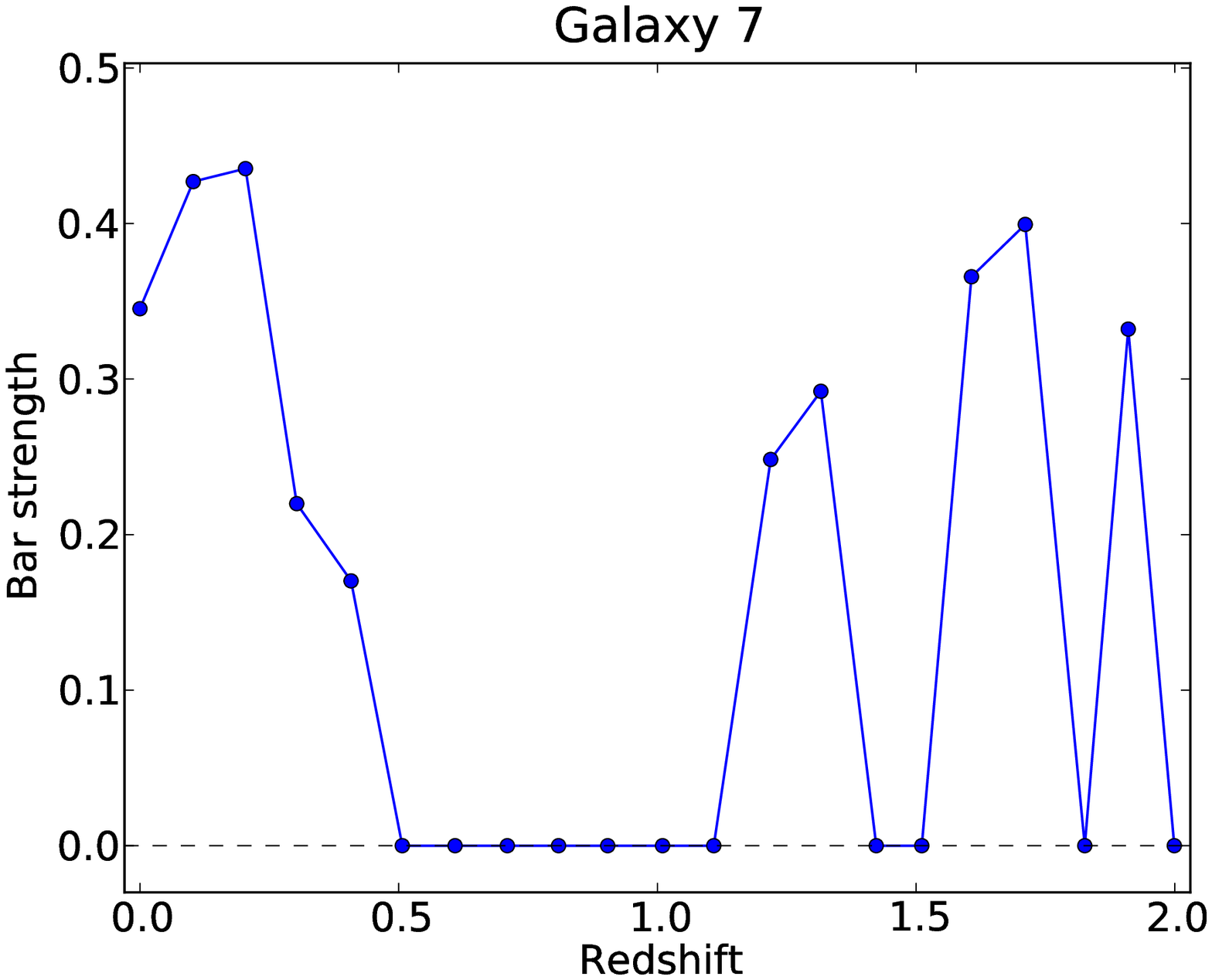}
\caption{\label{fig:early_bar} Examples of three simulated galaxies with an early bar formation ($1<z<2$). Galaxy 5 (top panel) hosts a bar that is destroyed 
spontaneously rapidly after its formation at $z\sim2$. A new bar reforms at low redshift ($z\sim0.2$). The bars developed at $z\gtrsim1$ in Galaxies 6 and 7 
(middle and bottom panels) are destroyed by mergers, but reform later. The early formed bars seem to be short-lived, although they can reform later. 
The high frequency at which the bar is destroyed and reformed between $z>1$ and $z<2$ in Galaxy 7 (bottom panel) is rather rare. Bars tend to undergo cycles 
of formation and destruction more commonly at higher redshifts, bars are also more rare at $z > 1$.}
\end{figure}

Galaxy~4 hosts a strong bar in the fiducial simulation, with $S>0.3$ from redshift $z\sim0.8$ down to $z\sim0$. When we shut down external accretion, the bar stays of a 
similar strength down to $z\sim 0.2$, and then weakens significantly.
In this case, external accretion is necessary to maintain a strong bar down to $z=0$.

Hence, bars that are formed after $z\sim0.8-1$ are mostly long-lived, often without requiring external accretion, which is probably explained by the fact that these 
bars form once the gas fraction is relatively low -- bar dissolution is easier to reach with high gas fractions. There are however cases where the bar evolution is 
influenced by external mass infall and late cosmological accretion below redshift one, including cases where the presence of a strong bar at $z=0$ is only achieved 
through external accretion.

\medskip

At the opposite, the (relatively rare) bars that are formed early ($1<z<2$) have a shorter lifetimes. They can either dissolve spontaneously, as is the case for the 
first galaxy shown on Figure~\ref{fig:early_bar}, or be destroyed by mergers, as is the case for the two other galaxies shown on this figure. The shorter bar lifetime 
at $z>1$ compared to $z<1$ can be explained by the higher gas fractions at these early epochs, making easier to dissolve a bar by gravitational torquing and 
mass concentration \citep{BCS05}. In these galaxies with early bars, the bar re-forms and eventually persist down to redshift zero, owing to late cosmological 
infall and build-up of a massive thin disk. 

Overall, it appears that external gas accretion is required to maintain low-redshift bars only in a limited fraction of the sample, but is generally required to 
re-form early bars at $z\sim1$, which tend to be intrinsically short-lived.

\section{The role of bars in (pseudo)-bulge growth}
\label{Sec:bars_vs_bulge}

In this Section, we address the impact of the bar in the growth of the bulge, since bars have been proposed to play an important role in the formation of pseudo-bulge 
by triggering the gas inflows to the central regions of galaxy \citep{2005MNRAS.358.1477A,2007ApJ...671..226H,2009ApJ...697..630F}. 

Here the identification of the bulge is made using GALFIT~\citep{Peng2002,Peng2010} for galaxies at $z=0$. Five galaxies (out of 33 analyzed so far) are removed from the 
study because their complex structure did not make it possible to achieve a satisfying and unique GALFIT decomposition. The majority of bulges have relatively low 
S\'{e}rsic indices indicating that they are pseudo-bulges rather than classical bulges (see \citealt{martig12} for a detailed study of the simulated galaxies in our sample).

We consider all stars that are present in the bulge of a given galaxy at $z=0$ and we compute the averaged normalized distribution of the redshift formation of these 
stars. For galaxies that host a bar this distribution is computed with respect to the moment of the apparition of the bar which is thus interpreted as the average 
redshift at which the bar forms. 

\begin{figure*}[t]
\includegraphics[width=\columnwidth]{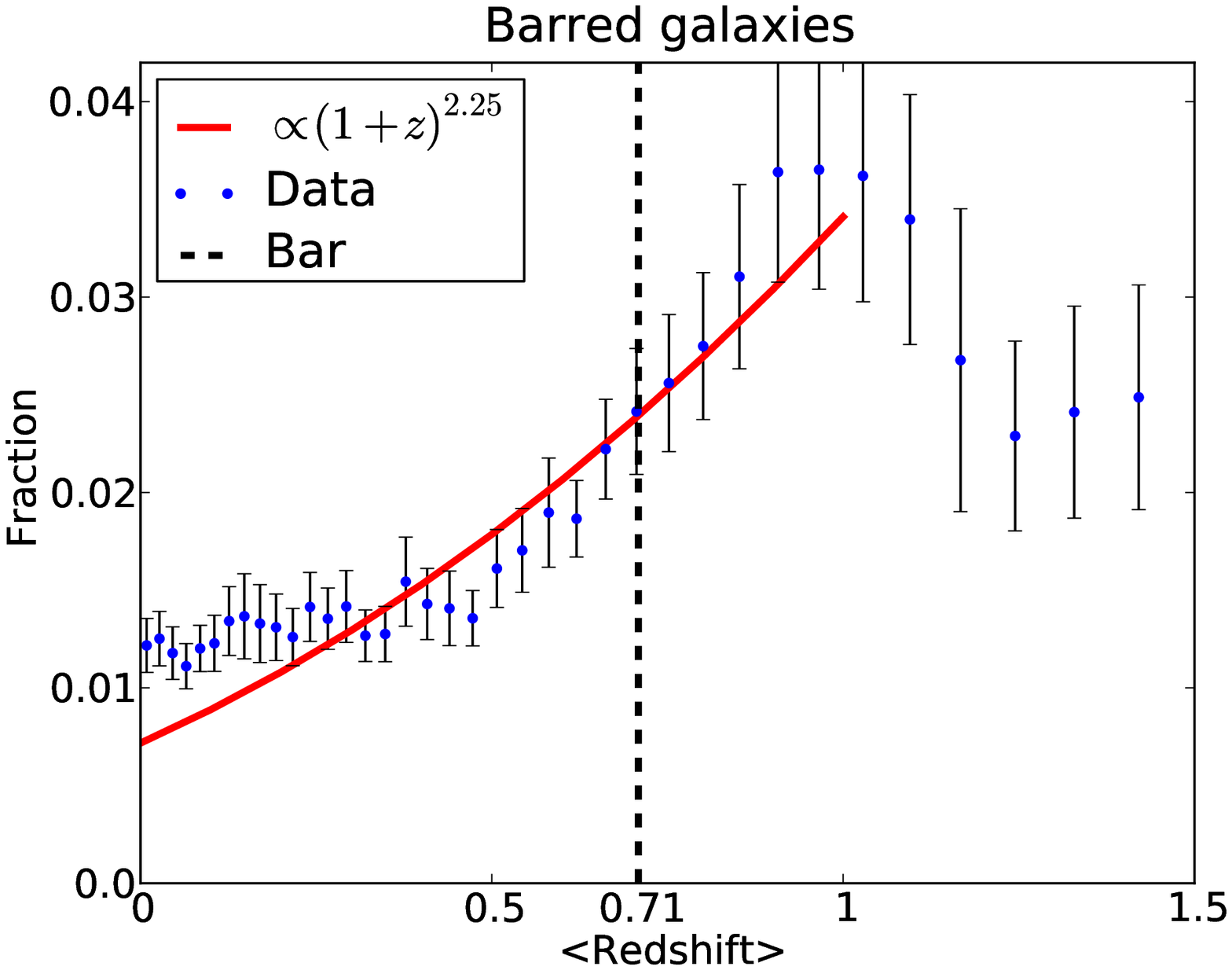}
\includegraphics[width=\columnwidth]{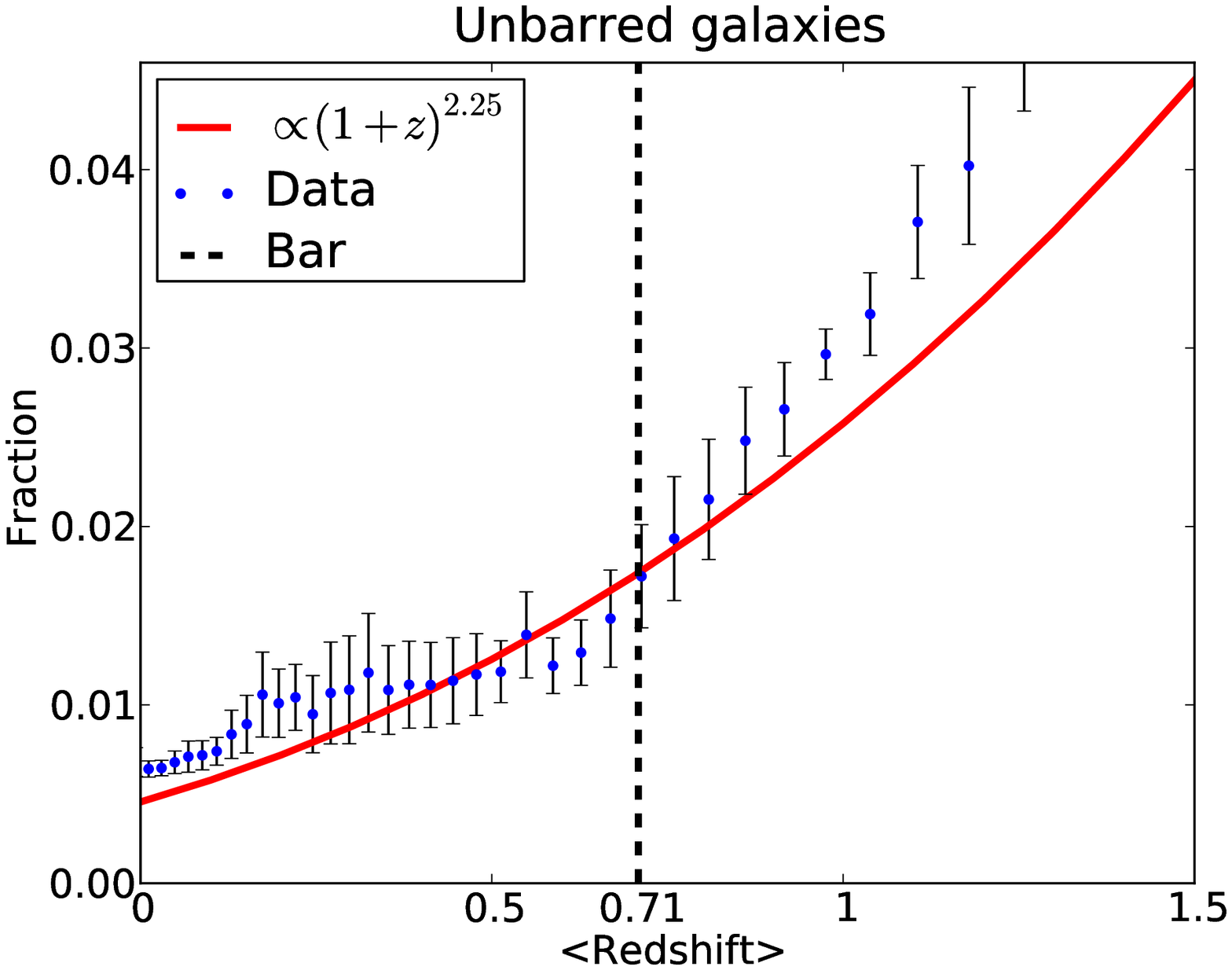}
\caption{\label{fig:fit_errors} Normalized bulge stellar age distribution. The two panels show the age distribution for stars in the bulge at $z=0$ for different types 
of galaxies (left panel: barred galaxies, right panel: unbarred galaxies). The fitted theoretical total mass accretion rate that scales as 
$(1+z)^{2.25}$ is shown too (the red line). The black dashed line represents the average time at which the bar is formed. All stars located to the left of this line 
are stars younger than the bar. We show this average bar formation redshift also for unbarred galaxies for comparison.}
\end{figure*}

In Figure~\ref{fig:fit_errors} we show the normalized age distribution of stars in the bulge for barred and unbarred galaxies (left and right panels, respectively), 
together with the theoretical mass accretion rate that scales with redshift as $(1+z)^{2.25}$ \citep{neistein06}. We note a slightly larger excess of stars in barred 
galaxies with respect to the infall rate as compared to that of unbarred galaxies.

The qualitative comparison suggests that the presence of a bar has an impact on the growth of the bulge. We quantify this influence in terms of the ratio of the area 
of the histogram over the region between the redshift $z=0$ and the average redshift of bar formation to the area over redshift range of $0\leq z\leq2$. We find 
$24\%$ for barred galaxies and $11\%$ for galaxies without a bar.

Even though it is difficult to quantify the effect of the bar on the bulge more precisely, both qualitative and quantitative study suggest that the bar plays a role 
in the growth of the bulge as cosmological accretion rate of baryons alone is not sufficient in explaining this over-abundance of star forming in the pseudo-bulge at 
late epochs.

%%%%%%%%%%%%%%%%%%%%%%%%%%%%%%%%%%%%%%%%%%%%%%%%%%%%%%%%%%%%%%%%%%%%%%%%%%%%%%%%%%%%%%%%%%%%%%%%%%%%%%%%%%%%%%%%%%%%%%%%%%%%%%%%%%%%%%%%%%%%%%%%%%%%%%%%%%%%%%%%%%
\section{Discussion: does the emergence of bars at $z \sim 1$ trace a new phase in spiral galaxy formation?}
\label{Sec:discussion}

In our models, bars emerge in present-day spirals at $z\sim 1$, being rare and only weak at higher redshift, and rapidly becoming ubiquitous at lower redshift. We 
suggested from our simulations that this corresponds to the transition between an early ``violent'' phase down to $z \simeq 0.8-1$ and a late ``secular'' phase at 
lower redshifts. Indeed, the morphology of the progenitors of today's spirals is rapidly evolving and uncorrelated with their final structure in the proposed ``violent''
phase at $z>1$ (see \citealt{martig12}), with mostly the thick disk and stellar spheroids forming in this phase, while after $z<1$ the thin disk grows and the structural parameters 
such as the bulge and disk fractions become well correlated with their final values at $z=0$. 

The influence of the dark matter halo can have additional effects on the evolution of bars (see e.g., \citealt{2006ApJ...648..807B,2010MNRAS.406.2386M}, for the 
effects of the shape of halo and \citealt{2007MNRAS.375..460W,2008ApJ...679..379S}, for the effects involving more general halo properties). 
As the dark matter halo evolves with redshift, it affects the evolution of the entire galaxy, 
in particular the bar, which in turn influences the halo itself. Bar formation can thus be reinforced or delayed depending on exact halo properties. Such effects 
should be resolved by our simulations, but the key epoch of bar formation appears to correspond mostly to the evolution of baryonic properties in our analysis.

\begin{figure*}[t]
\centering
\includegraphics[width=1.0\textwidth]{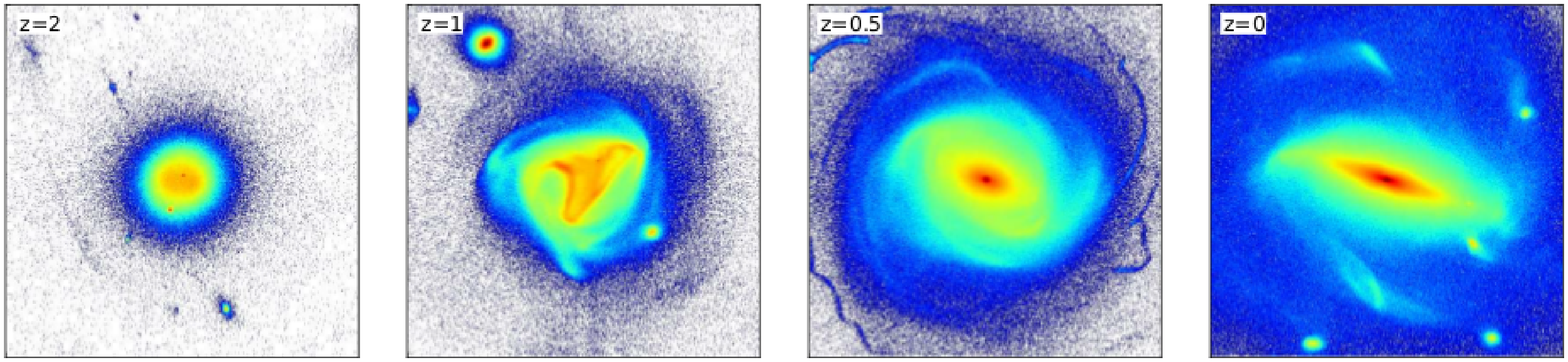}\\
\vskip 0.5cm
\includegraphics[width=1.0\textwidth]{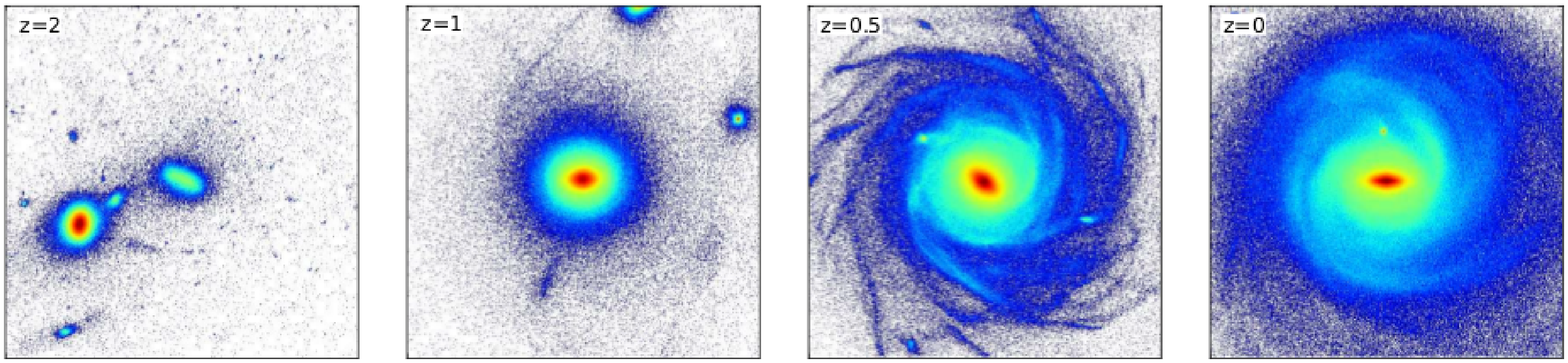}\\
\vskip 0.5cm
\includegraphics[width=1.0\textwidth]{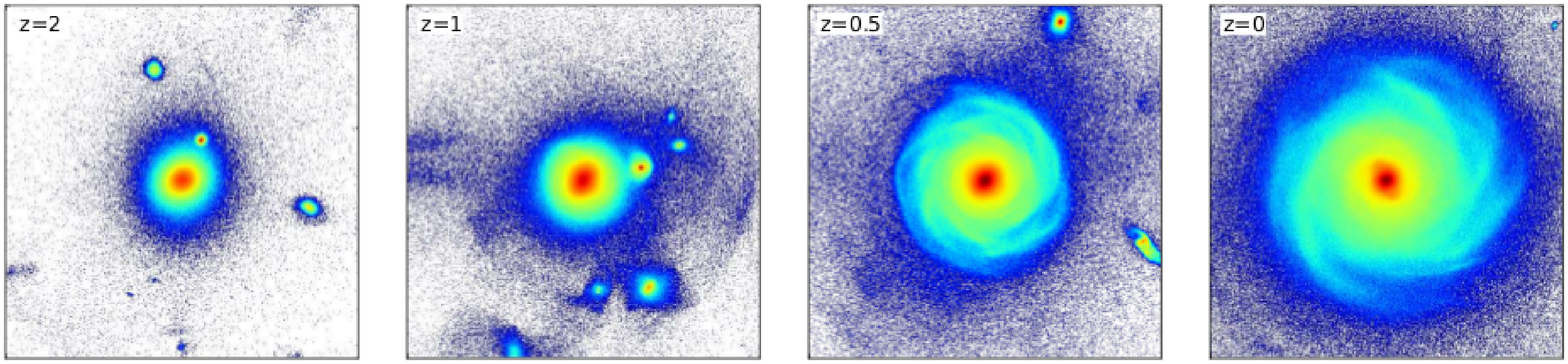}
\caption{\label{fig:images_z_evolution}
Examples of morphological evolution from $z=2$ down to $z=0$ for three simulated galaxies. Stellar density maps (face-on $50 \times 50$ kpc$^2$ projections) are shown 
for $z=2$, $1$, $0.5$ et $0$. Galaxies evolve rapidly during the ``violent'' phase at $z\gtrsim1$ when they frequently undergo phases of violent disk instabilities 
(top panels) and major mergers (middle panels). The morphology at $z=0$ and $z\gtrsim1$ are uncorrelated with early spheroids being possible 
progenitors of today's spirals (bottom panels). Once galaxies enter the late ``secular'' phase at $z<1$, their structural parameters become more tightly 
correlated with the final disk morphology. The color coding of the projected density maps is the same as in Figure~\ref{fig:lin_fit}.}
\end{figure*}

We show on Figure~\ref{fig:images_z_evolution} three representative examples at $z>1$: one, which is disk-dominated at $z>1$ but where violent disk instabilities 
(including giant clumps) destroy the early disk into a thick disk and spheroid, another one, which is spheroid-dominated with a major merger at $z\sim 2$, and a third 
one, which is also spheroid dominated with several minor mergers at $z\simeq1-2$. This is illustrative of the ``violent phase'' at $z>1$. Indeed, it is known
from other work that high-redshift disks have high gas fractions and are violently unstable with giant clumps and transient features, but do not frequently develop bars
\citep[e.g.,][]{BEE07, CDB10} and this instability destroys any thin disk that would have started to grow, while major mergers that can reform some disk components 
\citep[e.g.,][]{robertson06} but mostly convert disks into spheroids even in high-redshift conditions \citep{bournaud11}.

Hence, in the two examples on Figure~\ref{fig:images_z_evolution} and in the majority of our sample, no massive thin disk can stabilize before redshift one and develop 
a substantial bar. On the other hand, mergers with mass ratios larger than 5:1 are almost absent from our sample after $z=1$ and diffuse gas accretion occurs at much 
lower rate, and the rate of stellar bulge growth also drops. Thus a massive thin disk can form and start to dominate the mass distribution, as probed by the formation 
of spiral arms between $z=1$ and $z=0.5$ in the face-on images shown on Figure~\ref{fig:images_z_evolution}. This thin disk grows secularly down to $z=0$ 
and generally gets barred  by $z \approx 0.5$ -- strongly in the two first cases, weakly in the third one.

These three cases illustrate the transition between an early violent phase with frequent mergers and disk instabilities, and a late secular phase dominated by slower 
mass infall, and the fact that this transition occurs when the massive thin disk forms and is traced by the emergence of bars. Overall, the epoch of bar formation in our 
simulations probes the epoch at which spiral galaxies have formed the bulk of their disk, stellar halo and thick disk, and start to be dominated (in stellar mass) by 
their final thin disk with only slow (secular) evolution down to $z=0$.

\medskip

In observations, the bar fraction also decreases with increasing redshift, and although it could be probed in detail only up to $z \simeq 0.8$ it is in close agreement
 with our models so far. In the following we discuss whether the emergence of bars at $z\simeq 0.8-1$, if confirmed observationally, could also correspond to the 
epoch at which spiral galaxies acquire their final morphology and start being dominated by their thin rotating stellar disk. We are thus led to wonder whether a 
transition between a violent phase of galactic assembly, with rapid episodes of merging and violent instabilities building a thick disk and spheroids, could be followed
by a calm secular phase of thin disk growth and evolution, with the typical transition at $z\simeq 0.8-1$.

Chemical properties of disk galaxies could be consistent with such two-phase assembly of their components. Thick disks are ubiquitous around spiral galaxies 
\citep{dalcanton-bernstein02,seth05}, without star formation and no or almost no young stellar populations \citep{yoachim-dalcanton06, ibata09}. The Milky Way thick 
disk contains no or almost no star younger than 8~Gyr \citep{gilmore1985, reddy2006}. This is consistent with the occurrence of events disrupting any cold rotating disk
into a thick disk only at redshifts higher than $z\sim 1$, while the thin disk wouldn't have been significantly disrupted/thickened after this epoch. The enhanced 
$\alpha$ element abundances in the thick disk and central bulge \citep{lecureur07, zoccali07} suggest that their star formation occurred mostly in brief events (not 
longer than a few $10^8$~yr), and that the formation of stars that belong to the present-day thin disk occurred at later epochs with longer timescales 
\citep[see also][]{chiappini}. These brief events could have been mergers, as well as violent instabilities (giant clumps) in rapidly-accreting galaxies, that grow 
both a bulge and a thick disk \citep{BEM09} over short timescales, while a thin disk component can form and remain stable after these phases with lower gas infall 
rates.
 
\medskip

Direct searches for disks, through gas kinematics and/or optical and near-infrared spectroscopy, are also consistent with $z \simeq 1$ as the typical redshift for the 
emergence of modern thin spiral disks, along with their bars. Near-infrared spectroscopy of a $z\sim3$ sample of star-forming galaxies
(AMAZE and LSD, \citealt{maiolino10,2011A&A...528A..88G}), with masses typical for the progenitors of Milky Way-like galaxies and present-day spirals, reveals only a 
minor fraction of rotating disks, the majority of these galaxies being mergers or irregular systems dominated by high velocity dispersions. A similar survey of 
star-forming galaxies in a comparable mass range at a median redshift $z\simeq1.2$ (MASSIV, \citealt{2012A&A...539A..92E}) 
finds about 40-50\% of rotating disk galaxies. At $z\simeq 0.6-0.8$, the IMAGES survey finds a majority of rotating disk galaxies \citep{yang08} in a sample that 
still covers masses typical for the progenitors of today's spiral galaxies in the $10^{10-11}\,M_{\odot}$ mass range. This survey furthermore suggests that many of 
these $z\simeq 0.6-0.8$ galaxies have formed their disks only recently after undergoing violent events such as major mergers \citep{2009A&A...507.1313H} 
and that they will undergo only slow evolution of their global properties, such as their Tully-Fisher relation, down to $z=0$ \citep{puech10}.

Morphological studies are also consistent with the emergence of modern thin disks around redshift $z\sim1$. The Hubble Ultra Deep Field sample studied 
by \citet{elmegreen07,elmegreen09} at $z>1$ is dominated by irregular morphologies corresponding to major merger and interactions and ``clumpy'' unstable disks, 
which are typically forming thick disks, bulges and stellar haloes \citep{BEM09}, not thin spiral disks. Some cold spiral disks are found in this sample but their 
fraction is quite low before $z=1$. At $z\simeq0.7$, the situation is largely different as the fraction of clumpy irregular disks drops steadily and stable spiral 
disks rapidly become more numerous (still for progenitors of present-day $10^{10-11}\,M_{\odot}$ galaxies). For somewhat more massive galaxies, 
\citet{2007ApJS..172..434S} found that a large fraction of massive disks are in place around redshift one, but substantially fewer than at lower redshift.

\medskip

Hence, existing observations are consistent with our suggestion that the emergence of galactic bars at z$\simeq0.8-1$ traces the transition between an early 
``violent'' phase during which stars that belong to the modern thick disk, bulge and halo are formed in systems that do not have a permanent disk-dominated structure, 
and a late ``secular'' phase of thin disk growth and evolution with ubiquitous bars and limited pseudo-bulge growth. In this picture, the downsizing of bar 
formation (\citealt{2008ApJ...675.1141S} and Section~4.2) could correspond to the later termination of the violent phase and later disk stabilization in lower-mass 
galaxies. This would be consistent with the fact that both merging activity and violent disk instabilities should persist at lower redshift for lower mass galaxies 
\citep[][and references therein]{bournaud12}. It is also possible that bars grow more rapidly in more massive systems once their cold, thin disk is stabilized 
\citep{elmegreen-bars07}. The observed evolution of the bar fraction so far is consistent with our model, but further confirmation could be obtained by confirming the 
drop in the bar fraction at $z\sim1$ and above, with almost only weak bars (strength $\leq$ 0.2) being present at $z > 1$ for the mass range studied here.

%%%%%%%%%%%%%%%%%%%%%%%%%%%%%%%%%%%%%%%%%%%%%%%%%%

\section{Summary}
\label{Sec:summary}
We studied a sample of cosmological zoom-in simulations of 33 Milky Way-mass galaxies in field and loose group environments from $z=2$ down to $z=0$. 
The method used to determine the presence of a bar is based on the decomposition of the stellar surface density profiles into Fourier components. 
We also analyzed the disk/spheroid structure of the modeled galaxies using the S\'{e}rsic index of the surface density profile. 

Our main results are as follows:
\begin{enumerate}[i)]
 \item The \textit{total} bar fraction among spiral galaxies declines with increasing redshift. It drops from almost $90\%$ at $z=0$ to about $50\%$ at $z\simeq1$ and 
to almost zero at  $z\simeq2$. The fraction of \textit{observable} and \textit{strong} bars declines from about $70\%$ at $z=0$ to $10-20\%$ at $z\simeq1$ and to zero 
at $z=2$. This result holds for galaxies with mass range of $2\times 10^{9}-8\times 10^{10}$\,M$_{\sun}$ at $z\sim1$ and $4\times 10^{5}-1\times 10^{10}$\,M$_{\sun}$ at 
$z\sim2$, i.e. typical progenitors of Milky Way-like spirals. However, the bar fraction could remain higher at $z\sim2$ for more massive galaxies, if the downsizing of bar formation observed in our sample still holds for higher masses/redshifts.
 \item The epoch of bar formation traces the epoch of the emergence of final disk of spiral galaxies. This corresponds to the termination of an early ``violent'' phase 
at $z>1$, characterized by frequent mergers, violent disk instabilities and rapidly evolving structure, forming thick disks, bulges and stellar halos. It is followed by a ``secular'' phase at $z<0.8$, dominated by the slower growth and evolution of modern thin disks and limited bulge growth at late times. The $z=0.8-1.0$ transition epoch is for the mass of typical Milky Way progenitors, and tends to move to higher redshift for more massive systems.
 \item We find that there is only a minor contribution of bars in the late growth ($z<1$) of (pseudo-)bulges in spiral galaxies. This late growth is dominated by 
continued cosmic infall and minor mergers rather than by bars. 
 \item Finally, early bars (formed at $z>1$) are often short lived and may reform several times. Bars formed below $z\sim1$ are found to persist down to $z=0$, some of 
them being intrinsically short-lived but maintained by late cosmological gas infall.
\end{enumerate}

If confirmed observationally, the scarcity of significant bars at $z \geq 1$ would indicate, according to our models, that present day spirals and Milky Way-like 
galaxies have formed and stabilized their modern thin spiral disk only relatively late in their growth history, typically at $z \simeq 0.8-1$. At earlier times, 
they would be mostly forming their spheroidal components (bulges, halos) and thick disk, under the effect of both hierarchical merging and violent instabilities in 
rapidly-accreting systems. The continuation of this violent phase with mergers, rapid cold gas accretion and disk instabilities down to lower redshift in lower mass 
galaxies \citep[e.g.,][]{bournaud12} could then explain a ``downsizing''-like behaviour for bar formation.

\acknowledgments
We are grateful to Kartik Sheth, Fran\c{c}oise Combes and an anonymous referee for helpful comments. The simulations presented in the work were performed at the {\em Centre de Calcul Recherche et Technologie} of CEA under GENCI allocations 2011-GEN 2192 and 2012-GEN2192. We acknowledge financial support from CEA through a CFR grant (KK), from the EC through an ERC grant StG-257720 (FB, KK) and from the Australian government through a QEII Fellowship (MM).


\begin{thebibliography}{}
\bibitem[Abraham et al.(1999)]{1999MNRAS.308..569A} Abraham, R.~G., Merrifield, M.~R., Ellis, R.~S., Tanvir, N.~R., \& Brinchmann, J.\ 1999, \mnras, 308, 569 
\bibitem[Aguerri et al.(1998)]{aguerri1998} Aguerri, J.~A.~L., Beckman, J.~E., \& Prieto, M.\ 1998, \aj, 116, 2136 
\bibitem[Aguerri et al.(2009)]{2009A&A...495..491A} Aguerri, J.~A.~L., M{\'e}ndez-Abreu, J., \& Corsini, E.~M.\ 2009, \aap, 495, 491 
\bibitem[Athanassoula(2002)]{2002ApJ...569L..83A} Athanassoula, E.\ 2002, \apjl, 569, L83 
\bibitem[Athanassoula(2005)]{2005MNRAS.358.1477A} Athanassoula, E.\ 2005, \mnras, 358, 1477
\bibitem[Athanassoula et al.(2005)]{2005MNRAS.363..496A} Athanassoula, E., Lambert, J.~C., \& Dehnen, W.\ 2005, \mnras, 363, 496 
\bibitem[Barazza et al.(2008)]{2008ApJ...675.1194B} Barazza, F.~D., Jogee, S., \& Marinova, I.\ 2008, \apj, 675, 1194 
\bibitem[Barway et al.(2011)]{2011MNRAS.410L..18B} Barway, S., Wadadekar, Y., \& Kembhavi, A.~K.\ 2011, \mnras, 410, L18 
\bibitem[Berentzen et al.(2003)]{2003MNRAS.341..343B} Berentzen, I., Athanassoula, E., Heller, C.~H., \& Fricke, K.~J.\ 2003, \mnras, 341, 343 
\bibitem[Berentzen et al.(2004)]{2004MNRAS.347..220B} Berentzen, I., Athanassoula, E., Heller, C.~H., \& Fricke, K.~J.\ 2004, \mnras, 347, 220
\bibitem[Berentzen \& Shlosman(2006)]{2006ApJ...648..807B} Berentzen, I., \& Shlosman, I.\ 2006, \apj, 648, 807 
\bibitem[Block et al.(2002)]{block02} Block, D.~L., Bournaud, F., Combes, F., Puerari, I., \& Buta, R.\ 2002, \aap, 394, L35 
\bibitem[Bournaud \& Combes(2002)]{BC02} Bournaud, F., \& Combes, F.\ 2002, \aap, 392, 83
\bibitem[Bournaud \& Combes(2003)]{BC03} Bournaud, F., \& Combes, F.\ 2003, \aap, 401, 817 
\bibitem[Bournaud et al.(2005)]{BCS05} Bournaud, F., Combes, F., \& Semelin, B.\ 2005, \mnras, 364, L18 
\bibitem[Bournaud et al.(2007)]{BEE07} Bournaud, F., Elmegreen, B.~G., \& Elmegreen, D.~M.\ 2007, \apj, 670, 237 
\bibitem[Bournaud et al.(2009)]{BEM09} Bournaud, F., Elmegreen, B.~G., \& Martig, M.\ 2009, \apjl, 707, L1
\bibitem[Bournaud et al.(2011)]{bournaud11} Bournaud, F., Chapon, D., Teyssier, R., et al.\ 2011, \apj, 730, 4
\bibitem[Bournaud et al.(2012)]{bournaud12} Bournaud, F., Juneau, S., Le Floc'h, E., et al.\ 2012, ApJ submitted, arXiv:1111.0987 
\bibitem[Bureau \& Freeman(1999)]{1999AJ....118..126B} Bureau, M., \& Freeman, K.~C.\ 1999, \aj, 118, 126 
\bibitem[Burkert(2006)]{burkert06} Burkert, A.\ 2006, Comptes Rendus Physique, 7, 433
\bibitem[Ceverino et al.(2010)]{CDB10} Ceverino, D., Dekel, A., \& Bournaud, F.\ 2010, \mnras, 404, 2151 
\bibitem[Chiappini(2009)]{chiappini} Chiappini, C.\ 2009, IAU Symposium, 254, 191 
\bibitem[Combes \& Sanders(1981)]{combes-sanders} Combes, F., \& Sanders, R.~H.\ 1981, \aap, 96, 164
\bibitem[Combes \& Elmegreen(1993)]{combes-elmegreen} Combes, F., \& Elmegreen, B.~G.\ 1993, \aap, 271, 391 
\bibitem[Dalcanton \& Bernstein(2002)]{dalcanton-bernstein02} Dalcanton, J.~J., \& Bernstein, R.~A.\ 2002, \aj, 124, 1328 
\bibitem[Debattista \& Sellwood(2000)]{deb-sel2000} Debattista, V.~P., \& Sellwood, J.~A.\ 2000, \apj, 543, 704 
\bibitem[de Vaucouleurs et al.(1991)]{1991trcb.book.....D} de Vaucouleurs, G., de Vaucouleurs, A., Corwin, H.~G., Jr., et al.\ 1991, Volume 1-3, XII, 
2069 pp.~7 figs..~ Springer-Verlag Berlin Heidelberg New York
\bibitem[Elmegreen et al.(2004)]{2004ApJ...612..191E} Elmegreen, B.~G., Elmegreen, D.~M., \& Hirst, A.~C.\ 2004, \apj, 612, 191 
\bibitem[Elmegreen et al.(2007a)]{elmegreen07} Elmegreen, D.~M., Elmegreen, B.~G., Ravindranath, S., \& Coe, D.~A.\ 2007a, \apj, 658, 763 
\bibitem[Elmegreen et al.(2007b)]{elmegreen-bars07} Elmegreen, B.~G., Elmegreen, D.~M., Knapen, J.~H., et al.\ 2007b, \apjl, 670, L97
\bibitem[Elmegreen et al.(2009)]{elmegreen09} Elmegreen, B.~G., Elmegreen, D.~M., Fernandez, M.~X., \& Lemonias, J.~J.\ 2009, \apj, 692, 12
\bibitem[Epinat et al.(2012)]{2012A&A...539A..92E} Epinat, B., Tasca, L., Amram, P., et al.\ 2012, \aap, 539, A92
\bibitem[Eskridge et al.(2000)]{2000AJ....119..536E} Eskridge, P.~B., Frogel, J.~A., Pogge, R.~W., et al.\ 2000, \aj, 119, 536 
\bibitem[Fisher et al.(2009)]{2009ApJ...697..630F} Fisher, D.~B., Drory, N., \& Fabricius, M.~H.\ 2009, \apj, 697, 630
\bibitem[Friedli \& Benz(1993)]{1993A&A...268...65F} Friedli, D., \& Benz, W.\ 1993, \aap, 268, 65 
\bibitem[Gerin et al.(1990)]{gerin} Gerin, M., Combes, F., \& Athanassoula, E.\ 1990, \aap, 230, 37 
\bibitem[Gilmore et al.(1985)]{gilmore1985} Gilmore, G., Reid, N., \& Hewett, P.\ 1985, \mnras, 213, 257 
\bibitem[Gnerucci et al.(2011)]{2011A&A...528A..88G} Gnerucci, A., Marconi, A., Cresci, G., et al.\ 2011, \aap, 528, A88
\bibitem[Hammer et al.(2009)]{2009A&A...507.1313H} Hammer, F., Flores, H., Puech, M., et al.\ 2009, \aap, 507, 1313 
\bibitem[Hasan \& Norman(1990)]{norman} Hasan, H., \& Norman, C.\ 1990, \apj, 361, 69 
\bibitem[Hasan et al.(1993)]{hasan} Hasan, H., Pfenniger, D., \& Norman, C.\ 1993, \apj, 409, 91 
\bibitem[Heller et al.(2007)]{2007ApJ...671..226H} Heller, C.~H., Shlosman, I., \& Athanassoula, E.\ 2007, \apj, 671, 226
\bibitem[Hopkins et al.(2011)]{2011MNRAS.417..950H} Hopkins, P.~F., Quataert, E., \& Murray, N.\ 2011, \mnras, 417, 950
\bibitem[Ibata et al.(2009)]{ibata09} Ibata, R., Mouhcine, M., \& Rejkuba, M.\ 2009, \mnras, 395, 126 
\bibitem[Jogee et al.(2004)]{2004ApJ...615L.105J} Jogee, S., Barazza, F.~D., Rix, H.-W., et al.\ 2004, \apjl, 615, L105 
\bibitem[Jungwiert et al.(2001)]{2001A&A...376...85J} Jungwiert, B., Combes, F., \& Palou{\v s}, J.\ 2001, \aap, 376, 85
\bibitem[Kennicutt(1998)]{1998ApJ...498..541K} Kennicutt, R.~C., Jr.\ 1998, \apj, 498, 541 
\bibitem[Knapen et al.(2000)]{2000ApJ...529...93K} Knapen, J.~H., Shlosman, I., \& Peletier, R.~F.\ 2000, \apj, 529, 93 
\bibitem[Kormendy \& Kennicutt(2004)]{kormendy-kennicutt} Kormendy, J., \& Kennicutt, R.~C., Jr.\ 2004, \araa, 42, 603
\bibitem[Laine et al.(2002)]{2002ApJ...567...97L} Laine, S., Shlosman, I., Knapen, J.~H., \& Peletier, R.~F.\ 2002, \apj, 567, 97
\bibitem[Laurikainen et al.(2002)]{2002MNRAS.331..880L} Laurikainen, E., Salo, H., \& Rautiainen, P.\ 2002, \mnras, 331, 880
\bibitem[Laurikainen et al.(2004)]{2004ApJ...607..103L} Laurikainen, E., Salo, H., \& Buta, R.\ 2004, \apj, 607, 103
\bibitem[Lecureur et al.(2007)]{lecureur07} Lecureur, A., Hill, V., Zoccali, M., et al.\ 2007, \aap, 465, 799 
\bibitem[Lee et al.(2012)]{2012ApJ...745..125L} Lee, G.-H., Park, C., Lee, M.~G., \& Choi, Y.-Y.\ 2012, \apj, 745, 125 
\bibitem[Machado \& Athanassoula(2010)]{2010MNRAS.406.2386M} Machado, R.~E.~G., \& Athanassoula, E.\ 2010, \mnras, 406, 2386
\bibitem[Maiolino et al.(2010)]{maiolino10} Maiolino, R., Mannucci, F., Cresci, G., et al.\ 2010, The Messenger, 142, 36
\bibitem[Martig \& Bournaud(2010)]{2010ApJ...714L.275M} Martig, M., \& Bournaud, F.\ 2010, \apjl, 714, L275 
\bibitem[Martig et al.(2009)]{martig09} Martig, M., Bournaud, F., Teyssier, R., \& Dekel, A.\ 2009, \apj, 707, 250 
\bibitem[Martig et al.(2012)]{martig12} Martig, M., Bournaud, F., Croton, D.~J., Dekel, A., \& Teyssier, R.\ 2012, ApJ submitted, arXiv:1201.1079
\bibitem[Martinez-Valpuesta et al.(2006)]{martinez-v06} Martinez-Valpuesta, I., Shlosman, I., \& Heller, C.\ 2006, \apj, 637, 214
\bibitem[Men{\'e}ndez-Delmestre et al.(2007)]{2007ApJ...657..790M} Men{\'e}ndez-Delmestre, K., Sheth, K., Schinnerer, E., Jarrett, T.~H., \& Scoville, N.~Z.\ 2007, \apj, 657, 790
\bibitem[Miwa \& Noguchi(1998)]{1998ApJ...499..149M} Miwa, T., \& Noguchi, M.\ 1998, \apj, 499, 149
\bibitem[Neistein et al.(2006)]{neistein06} Neistein, E., van den Bosch, F.~C., \& Dekel, A.\ 2006, \mnras, 372, 933
\bibitem[Peng et al.(2002)]{Peng2002} Peng, C.~Y., Ho, L.~C., Impey, C.~D., \& Rix, H.-W.\ 2002, \aj, 124, 266 
\bibitem[Peng et al.(2010)]{Peng2010} Peng, C.~Y., Ho, L.~C., Impey, C.~D., \& Rix, H.-W.\ 2010, \aj, 139, 2097 
\bibitem[Pfenniger \& Friedli(1991)]{pfenniger} Pfenniger, D., \& Friedli, D.\ 1991, \aap, 252, 75 
\bibitem[Puech et al.(2010)]{puech10} Puech, M., Hammer, F., Flores, H., et al.\ 2010, \aap, 510, A68 
\bibitem[Reddy et al.(2006)]{reddy2006} Reddy, B.~E., Lambert, D.~L., \& Allende Prieto, C.\ 2006, \mnras, 367, 1329 
\bibitem[Robertson et al.(2006)]{robertson06} Robertson, B., Bullock, J.~S., Cox, T.~J., et al.\ 2006, \apj, 645, 986 
\bibitem[Sargent et al.(2007)]{2007ApJS..172..434S} Sargent, M.~T., Carollo, C.~M., Lilly, S.~J., et al.\ 2007, \apjs, 172, 434 
\bibitem[Sellwood(2008)]{2008ApJ...679..379S} Sellwood, J.~A.\ 2008, \apj, 679, 379 
\bibitem[S{\'e}rsic(1963)]{1963BAAA....6...41S} S{\'e}rsic, J.~L.\ 1963, Boletin de la Asociacion Argentina de Astronomia La Plata Argentina, 6, 41 
\bibitem[Seth et al.(2005)]{seth05} Seth, A.~C., Dalcanton, J.~J., \& de Jong, R.~S.\ 2005, \aj, 130, 1574 
\bibitem[Shen \& Sellwood(2004)]{shen-sellwood} Shen, J., \& Sellwood, J.~A.\ 2004, \apj, 604, 614 
\bibitem[Sheth et al.(2003)]{2003ApJ...592L..13S} Sheth, K., Regan, M.~W., Scoville, N.~Z., \& Strubbe, L.~E.\ 2003, \apjl, 592, L13 
\bibitem[Sheth et al.(2008)]{2008ApJ...675.1141S} Sheth, K., Elmegreen, D.~M., Elmegreen, B.~G., et al.\ 2008, \apj, 675, 1141 
\bibitem[Teyssier(2002)]{teyssier02} Teyssier, R.\ 2002, \aap, 385, 337 
\bibitem[Teyssier et al.(2010)]{2010ApJ...720L.149T} Teyssier, R., Chapon, D., \& Bournaud, F.\ 2010, \apjl, 720, L149
\bibitem[Toomre(1963)]{toomre63} Toomre, A.\ 1963, \apj, 138, 385 
\bibitem[van den Bergh(2002)]{vdbergh} van den Bergh, S.\ 2002, \aj, 124, 782 
\bibitem[Weinberg(1985)]{1985MNRAS.213..451W} Weinberg, M.~D.\ 1985, \mnras, 213, 451 
\bibitem[Weinberg \& Katz(2007)]{2007MNRAS.375..460W} Weinberg, M.~D., \& Katz, N.\ 2007, \mnras, 375, 460 
\bibitem[Whyte et al.(2002)]{whyte02} Whyte, L.~F., Abraham, R.~G., Merrifield, M.~R., et al.\ 2002, \mnras, 336, 1281
\bibitem[Yang et al.(2008)]{yang08} Yang, Y., Flores, H., Hammer, F., et al.\ 2008, \aap, 477, 789 
\bibitem[Yoachim \& Dalcanton(2006)]{yoachim-dalcanton06} Yoachim, P., \& Dalcanton, J.~J.\ 2006, \aj, 131, 226 
\bibitem[Zoccali et al.(2007)]{zoccali07} Zoccali, M., Lecureur, A., Barbuy, B., et al.\ 2007, IAU Symposium, 241, 73
\end{thebibliography}
\end{document}